\documentclass[11pt]{article}
\pdfoutput=1
\usepackage{jheppub}

\usepackage{amssymb,amsmath}
\usepackage{amsfonts}
\usepackage{mathtools}
\allowdisplaybreaks

\usepackage{physics}
\usepackage{cancel}
\usepackage{enumitem}
\usepackage[usenames,dvipsnames]{xcolor}
\usepackage{bm}
\usepackage{bbm}
\usepackage{slashed}
\usepackage{mathrsfs}

\let\originalleft\left
\let\originalright\right
\renewcommand{\left}{\mathopen{}\mathclose\bgroup\originalleft}
\renewcommand{\right}{\aftergroup\egroup\originalright}

\makeatletter
\newenvironment{equations}[1][]{\subequations\ifx\relax#1\relax\else\label{#1}\fi\align\ignorespaces}{\endalign\ignorespacesafterend\endsubequations}
\def\@spliteq#1{\begin{equation}\begin{split}#1\end{split}\end{equation}}
\def\splitequation{\collect@body\@spliteq}

\makeatother

\newcommand{\diff}{\mathrm{d}}
\newcommand{\de}{\partial}
\newcommand{\vepsilon}{\varepsilon}

\newcommand{\vphi}{\varphi}
\newcommand{\vrho}{\varrho}

\makeatletter
\@addtoreset{equation}{section}

\makeatother

\newcommand{\nn}{\nonumber}

\makeatletter
\gdef\@fpheader{}
\makeatother

\title{Anomalies in Covariant Fracton Theories}

\author[a,b]{Davide Rovere}

\affiliation[a]{Dipartimento di Fisica e Astronomia ``Galileo Galilei'', Universit\`a di Padova, Via F. Marzolo 8, 35131 Padua, Italy}
\affiliation[b]{INFN, Sezione di Padova, Padua, Italy}

\emailAdd{davide.rovere@studenti.unipd.it}

\abstract{Covariant (Lorentz invariant) fracton matter, minimally coupled and charged under a symmetric rank two gauge tensor, is considered. The gauge transformations correspond to linearized longitudinal diffeomorphisms. Consistent possible anomalies are computed using the \textsc{brst} cohomology method. They depend only on the gauge field, treated as a background field, and on the gauge parameter, promoted to an anticommuting scalar ghost field. The problem is phrased in terms of polyforms, whose total degree is the sum of the form degree and of the ghost number. The most general anomaly in two dimensions and in four dimensions is computed and an anomaly in arbitrary dimensions is individuated. In conclusion, it is shown that a simple higher-derivative scalar field theory is an example of covariant fracton matter.}

\begin{document}
\maketitle

\section{Introduction}

The main aim of this paper is to compute possible anomalies in covariant fracton matter theories, using the \textsc{brst} cohomology approach. Fracton phases have recently attracted considerable interest. In a two-dimensional lattice, a fracton phase can be defined as excitations with restricted mobility on a subspace of the whole space and with infinite degenerate ground state in the continuum limit \cite{Pretko:2017xar, Seiberg:2020bhn, Burnell:2021reh}.\footnote{The use of the word ``fracton" is borrowed from condensed matter physics. The first appearance in that context, to best knowledge of the author, goes back to 1982 \cite{Alexander:1982}, where it means ``quantized vibrational states on a fractal". Then, fractons were rediscovered (to the best knowledge of the author) in \cite{Vijay:2015mka}, where the same word is used to refer to excitations in spin models, characterized by reduced mobility, in the sense that they are free to propagate only in submanifolds of the lattice (``The fundamental excitations are termed `fractons' as they behave as fractions of a mobile particle"). Curiously, the word ``fracton" appeared in High Energy Physics in 1981 in \cite{Khlopov:1981wm} to denote colorless mesons or baryons with fractional electric charge (the author thanks M. Yu. Khlopov for having pointed this out).} The restricted mobility can be thought in terms of the conservation of the dipole momentum. Fractons are charged with respect to a $\text{U}(1)$ gauge field, whose temporal part is a $\text{U}(1)$ gauge field $A_0$, whereas its spatial part is a $\text{U}(1)$ rank two  symmetric tensor $A_{ij}$, $i,j=x,y$, whose infinitesimal gauge redundancy involves a second derivative:
\begin{equation}
\delta_\lambda\,A_0 = \de_0\,\lambda, \quad
\delta_\lambda\,A_{ij} = \de_i\,\de_j\,\lambda,
\end{equation}
where $\de_0$ is the time derivative and $\lambda$ is the gauge parameter. In this sense, fracton matter coupled to the gauge field is a sort of higher rank electrodynamics. The corresponding Noether currents $J^0, J^{ij}$ satisfy
\begin{equation}\label{NoCovConsEq}
\de_0\,J^0 + \de_i\,\de_j\,J^{ij} = 0,
\end{equation}
which implies not only the conservation of the charge $\int\diff^2x\,J^0$, but also the conservation of the dipole momentum $\int\diff^2x\,x^i\,J^0$. 

If $A_0, A_{ij}$ are replaced by a single covariant rank two symmetric tensor $h_{\mu\nu}$, a \emph{covariant fracton theory} comes up. Throughout the paper by ``covariant" we mean covariance with respect to the Lorentz group. This model was introduced for the first time in \cite{Blasi:2022mbl}, where its classical properties are analyzed.  In the covariant theory, the gauge redundancy reads
\begin{equation}\label{GaugeTransformation}
\delta_\lambda\,h_{\mu\nu} = \de_\mu\,\de_\nu\,\lambda.
\end{equation}

In a condensed matter context the relativistic invariance is not important, and the direct physical relevance of the covariant fracton models is not clear until now. Nevertheless, in \cite{Bertolini:2022ijb} it is shown that a covariant formulation allows one to derive more easily and to elucidate the main properties of fractons, in the same way as in ordinary covariant electrodynamics. Moreover, in \cite{Bertolini:2023juh} the formal relation with linearized Einstein gravity is investigated: $h_{\mu\nu}$ can be compared with the linear perturbation of the flat background metric, according to $g_{\mu\nu} = \eta_{\mu\nu} + \kappa^\alpha\,h_{\mu\nu}$, where $\kappa$ is the gravitational constant and $\alpha$ is an integer, such that $\kappa^\alpha\,h_{\mu\nu}$ is dimensionless. The gauge redundancy resembles the transformation of the perturbation of the metric under infinitesimal linearized diffeomorphisms
$\delta_\xi\,h_{\mu\nu} = \de_\mu\,\xi_\nu + \de_\nu\,\xi_\mu$, if one sets $\xi_{\mu} = \tfrac{1}{2}\,\de_\mu\,\lambda$. In this sense, the kinetic action of the fracton gauge field $h_{\mu\nu}$ is fixed only by the invariance under ``longitudinal diffeomorphisms". This brings to the usual linearized Einstein-Hilbert action plus another piece, which breaks the invariance under the whole diffeomorphism group, being invariant under the longitudinal diffeomorphisms only.

A covariant fracton matter theory is expected to be invariant under a global transformation, whose  Noether current is a covariant rank two symmetric tensor $J^{\mu\nu}$. It satisfies on shell a ``second derivative conservation law":
\begin{equation}
\de_\mu\,\de_\nu\,J^{\mu\nu} = 0,
\end{equation}
which is the covariant version of \eqref{NoCovConsEq}. $h_{\mu\nu}$ should gauge the global symmetry, promoting it to a local symmetry. The minimal coupling term $h_{\mu\nu}\,J^{\mu\nu}$ should dictate the linear dependence on the fracton gauge field in the locally invariant theory. The inclusion of the most general action quadratic in the derivatives of $h_{\mu\nu}$, invariant under the gauge transformation \eqref{GaugeTransformation}, brings to treat the gauge field dynamically.

The quantization of the fracton matter theory can be performed in the path integral formulation, leaving the fracton gauge field as a nonintegrated background field. One may ask if the global symmetry, enjoyed by the fracton matter theory at classical level, is still a symmetry in the quantum theory, that is, if the theory is afflicted or not by anomalies. If an anomaly exists, the conservation law breaks down:
\begin{equation}
\de_\mu\,\de_\mu\,\langle J^{\mu\nu}\rangle = \mathcal{A},
\end{equation}
where $\langle\dots\rangle$ is the quantum mean value and $\mathcal{A}$ is an anomaly. Here we are dealing with possible anomalies of a global symmetry, since the fracton gauge field is not dynamical. Thus, such an anomaly does not bring to inconsistencies. 

The computation of possible anomalies can be conveniently recasted in a cohomology problem, by means of the \textsc{brst} formalism \cite{Becchi:1974md, Becchi:1974xu, Becchi:1975nq, Tyutin:1975qk}. In this approach, the $d$-dimensional possible anomalies, which cannot be removed by local counterterms, are identified with the classes of the \textsc{brst} cohomology mod $\diff$ on $d$-forms of ghost number one:
\begin{equation}
s\,\omega_{1}^{(d)} = -\diff \omega_2^{(d-1)},
\end{equation} 
where $s$ is the \textsc{brst} operator, which generates the infinitesimal gauge transformations, $\diff$ is the de Rham differential and $\omega_g^{(p)}$ is a $p$-form of ghost number $g$. The ghost is the gauge parameter $\lambda$, promoted to an anticommuting scalar field, in such a way that the \textsc{brst} operator is nilpotent:
\begin{equations}
& s\,\lambda = 0,\\
& s\,h_{\mu\nu} = \de_\mu\,\de_\nu\,\lambda.
\end{equations} 

Since we are interested in the $s$ cohomology mod $\diff$, a descent of equations is considered. In Sections \ref{BRSTcoho} and \ref{cohoClasses}, we analyze the \textsc{brst} cohomology in fracton theories, in arbitrary spacetime dimensions, showing that there is no anomaly in odd dimensions. 

In order to deal with the problem of determining possible anomalies in arbitrary even dimensions, we work out a formulation of the model in terms of \emph{polyforms} (Section \ref{polyforms}). Stora and Zumino realized that the computation of anomalies in the Yang-Mills case and in the gravitational one is elucidated if one studies the cohomology of the operator
\begin{equation}
\delta = \diff + s,
\end{equation} 
which is equivalent to the $s$ cohomology mod $\diff$ \cite{Stora:1976LM, Stora:1984, Zumino:1983ew, Manes:1985df, Langouche:1984gn}. $\delta$ has to be thought as a nilpotent differential operator acting on polyforms \cite{Thierry-Mieg:1979fvq}. A polyform formally sums differential forms with various form degree (the grading defined by $\diff$) and ghost number (the grading defined by $s$), in such a way that the sum of the two degrees is the same for each term. This degree is called \emph{total degree} of the polyform. The Stora-Zumino approach turns out to be equivalent to the perhaps more familiar ``two-step descent procedure", which defines the so-called anomalous polynomial \cite{Zumino:1983rz}, but it has the advantage of not requiring additional spacetime dimensions. (See \cite{Imbimbo:2023sph} for a comparison of the two methods.) 

In Section \ref{SomeSolutions}, we compute the most general anomaly in two dimensions and in four dimensions, and we find an anomaly in arbitrary even dimensions. We argue that many other structures are allowed in higher dimensions, and we produce other anomalies in six dimensions in Section \ref{More6d}. In Section \ref{Weyl}, we also comment on the similarity with the well known type \textsc{a} Weyl anomaly in conformal gravity \cite{Deser:1993yx}.

It is important to stress that the individuation of a \textsc{brst} cohomology class does not necessarily lead to an anomaly in a physical field theory. If a suitable regularization scheme exists for a given theory, preserving the symmetry, there is no anomaly at all. In other terms, the cohomology approach allows one to compute the geometrical part of a possible anomaly, which is universal. However, the multiplicative coefficient depends on the dynamics of the specific physical theory, and it could vanish, according to the matter content.

In this paper we address only the issue of classification of the cohomology classes. The interesting question of the computation of the dynamical coefficient relies on the formulation of fracton matter theories. In Section \ref{Matter}, we introduce the most simple covariant fracton matter theory. It is a higher derivative theory of a scalar field. Nevertheless, being a St\"uckelberg version for the mass term for the fracton gauge field, it does not produce loop contributions.

\section{Covariant fracton gauge theories}\label{CovariantFractoGaugeTheories}

Consider a matter theory in a $d$-dimensional spacetime, whose matter fields $\vphi^i$ transform in some representation of the $d$-dimensional Lorentz group, and whose Lorentz invariant action $S[\vphi^i]$ is invariant under a global transformation too. Suppose this symmetry to be promoted from global to local by a rank two symmetric gauge field $h_{\mu\nu}$, $\mu,\nu=0,1,\dots,d-1$, with gauge transformation
\begin{equation}
\delta_\lambda\,h_{\mu\nu} = \de_\mu\,\de_\nu\,\lambda,
\end{equation}
with $\lambda$ gauge parameter. The local invariance is realized by an improved action $\hat{S}[\vphi^i,h_{\mu\nu}]$, which is the same as the matter action if the gauge field is set to zero: $\hat{S}[\vphi^i,0]=S[\vphi^i]$. 

$h_{\mu\nu}$ is called \emph{fracton gauge field}, since it is (the covariant version of) the gauge field first introduced in fracton matter models \cite{Pretko:2017xar, Gromov:2018nbv, Seiberg:2020bhn}. These models, which break the Lorentz invariance, can be viewed as the low energy continuum limit of some lattice models, whose excitations, called \emph{fractons}, are mainly characterized by reduced mobility on subspaces of the spacetime and by infinite degeneracy of the ground state. Here we deal with possible covariantized version of such fracton theories. 

The restricted mobility of fracton matter is a consequence of the conservation of the dipole momentum. If $Q$ is the conserved charge
\begin{equation}
Q = \int\diff^{d-1}x\,J^0, \quad \text{such that}\;\;\frac{\diff Q}{\diff t} = 0,
\end{equation}
for some time component $J^0$ of a covariant current, the corresponding conserved dipole momentum is
\begin{equation}
Q^k = \int\diff^{d-1} x\,x^k\,J^0 \quad\text{such that}\;\; \frac{\diff Q^k}{\diff t} = 0,
\end{equation}
where $k = 1,\dots,d-1$ runs over the space dimensions. These conservation laws are covariantly implied by supposing a symmetric current $J^{\mu\nu}$ to exist, such that
\begin{equation}\label{WeakConservation}
\de_\mu\,\de_\nu\,J^{\mu\nu} = 0.
\end{equation}
Indeed, this equation reads
\begin{equation}\label{WeakConservation2}
\de_0^2\,J^{00} + 2\,\de_0\,\de_i\,J^{0i} + \de_i\,\de_j\,J^{ij} = 0,
\end{equation}
$i,j=0,\dots,d-1$. Integrating in the space, the last two terms vanish if there is no boundary, and the first one brings to a conserved quantity
\begin{equation}
\int\diff^{d-1}x\,\de_0\,J^{00} = \text{cost,}
\end{equation}
identified with $Q$, if $J^0 = \de_0\,J^{00}$. Moreover, multiplying the equation \eqref{WeakConservation2} by $x^k$ and integrating in the space,
\begin{align}
0 &= \de_0^2\int\diff^{d-1}x\,x^k\,J^{00} + 2\,\de_0\int\diff^{d-1}x\,x^k\,\de_i\,J^{0i} + \int\diff^{d-1}\,x^k\,\de_i\,\de_j\,J^{ij} = \nn\\
& = \de_0^2\int\diff^{d-1}x\,x^k\,J^{00} - 2\,\de_0\int\diff^{d-1}x\,\delta^k_i\,J^{0i} - \int\diff^{d-1}x\,\delta^k_i\,\de_j\,J^{ij} = \nn\\
& = \de_0^2\int\diff^{d-1}x\,x^k\,J^{00} - 2\,\de_0\int\diff^{d-1}x\,J^{0k} = \de_0\,\left(\int\diff^{d-1}x\,x^k\,\de_0\,J^{00} - 2\int\diff^{d-1}x\,J^{0k}\right).
\end{align}
If the last term drops out (for example, if $J^{0k}$ is a total spatial derivative -- in Section \ref{Matter} a model with this property is considered), then
\begin{equation}
\int\diff^{d-1}x\,x^k\,\de_0\,J^{00} = \text{cost},
\end{equation}
so that it can be identified with the dipole momentum $Q^k$.

In \cite{Bertolini:2022ijb, Bertolini:2023juh} the ``fracton field strength" is defined as\footnote{This tensor was introduced for the first time in \cite{Wu:1988py}, where the gauge \eqref{GaugeTransformation} transformation also appears.}
\begin{equation}
F_{\mu\nu\vrho} := \de_{\mu}\,h_{\nu\vrho} + \de_\nu\,h_{\mu\vrho} - 2\,\de_{\vrho}\,h_{\mu\nu},
\end{equation}
which is symmetric in the first two indices. It satisfies a cyclic identity and a Bianchi identity, and it is gauge invariant:
\begin{equations}
& F_{\mu\nu\vrho} + F_{\nu\vrho\mu} + F_{\vrho\mu\nu} = 0,\\
& \vepsilon^{\sigma\nu\vrho}\,\de_\sigma\,F_{\mu\nu\vrho}=0, \label{cyclicF}\\
& \delta_\lambda\,F_{\mu\nu\vrho} = 0.\label{Bianchi}
\end{equations}
Consider for definiteness the three-dimensional case, which is the first nontrivial one. The first and the second properties imply $F_{\mu\nu\vrho}$ to sit in the representation $\mathbf{1}\oplus\mathbf{2}$ of the three-dimensional Lorentz group, contained in $\mathbf{1}\otimes \mathbf{1} \otimes \mathbf{1}$.\footnote{In general, a tensor $T_{\mu\nu\alpha}$ transforms in the representation $\mathbf{1}\otimes\mathbf{1}\otimes\mathbf{1}$ of the Lorentz group in three dimensions. The factorization in irreducible spin representations is
\begin{equation}
T_{\mu\nu\alpha} \rightarrow \mathbf{1}\otimes\mathbf{1}\otimes\mathbf{1} = 
\mathbf{1}\oplus(\mathbf{0}\oplus\mathbf{1}\oplus\mathbf{2})\oplus(\mathbf{1}\oplus\mathbf{2}\oplus\mathbf{3}).
\end{equation}
The unique scalar is $\vepsilon^{\mu\nu\alpha}\,T_{\mu\nu\alpha}$, which is dual to the totally antisymmetric part $(\mathbf{1}\otimes\mathbf{1}\otimes\mathbf{1})_A$, given by $T_{[\mu\nu\vrho]}$. The three vectors $\mathbf{1}$ are the possible traces $T_\mu{}^\mu{}_\alpha$, $T_{\mu\nu}{}^\mu$ and $T_{\mu\nu}{}^\nu$. Let $\tilde{T}_{\mu\nu\alpha}$ the symmetric part in the first two indices $\frac{1}{2}\,T_{(\mu\nu)\alpha}$, which therefore transforms in the representation $(\mathbf{0}\oplus\mathbf{2})\otimes\mathbf{1}$:
\begin{equation}
\tilde{T}_{\mu\nu\alpha} \rightarrow (\mathbf{0}\oplus\mathbf{2})\otimes\mathbf{1} = \mathbf{1}\oplus(\mathbf{1}\oplus\mathbf{2}\oplus\mathbf{3}).
\end{equation}
The totally symmetric part of $\tilde{T}_{\mu\nu\alpha}$ is twice the cyclic combination $\tilde{T}_{\mu\nu\alpha}+\tilde{T}_{\nu\alpha\mu}+\tilde{T}_{\alpha\mu\nu}$. The totally symmetric part of $\mathbf{1}\otimes\mathbf{1}\otimes\mathbf{1}$ is $\mathbf{1}\oplus\mathbf{3}$. Therefore, imposing the cyclic identity 
\begin{equation}
\tilde{T}_{\mu\nu\alpha}+\tilde{T}_{\nu\alpha\mu}+\tilde{T}_{\alpha\mu\nu} = 0
\end{equation}
corresponds to select the representation $\mathbf{1}\oplus\mathbf{2}$ within $\tilde{T}_{\mu\nu\alpha}$. Thus, $\mathbf{1}\oplus\mathbf{2}$ is the representation associated to the fracton curvature $F_{\mu\nu\alpha}$. The vector component is the unique independent trace, say, $F^\mu{}_{\mu\alpha}$; the $\mathbf{2}$ is the traceless part, which turns out to be equal to
\begin{equation}
F_{\mu\nu\alpha} - \tfrac{1}{4}\,(2\,\eta_{\mu\nu}\,F^\lambda{}_{\lambda\alpha} - \eta_{\mu\alpha}\,F^\lambda{}_{\lambda\nu} - \eta_{\nu\alpha}\,F^\lambda{}_{\lambda\mu}).
\end{equation}} The $\mathbf{1}$ is the unique trace of $F_{\mu\nu\vrho}$. The trace is unique since the cyclic identity implies $F^\mu{}_{\nu\mu} = -\tfrac{1}{2}\,F^\mu{}_{\mu\nu}$. The representation content of $F_{\mu\nu\vrho}$ suggests it to be expressed in terms of a rank two traceless tensor. This is obtained by introducing the dual of $F_{\mu\nu\vrho}$:
\begin{equation}\label{E3d}
e_{\mu\nu} :=\tfrac{1}{3}\,\vepsilon_\mu{}^{\alpha\beta}\,F_{\nu\alpha\beta} = \vepsilon_\mu{}^{\alpha\beta}\,\de_\alpha\,h_{\beta\nu}\quad\text{in}\;d=3\;\text{dimensions}.
\end{equation}
The cyclic identity and the Bianchi one correspond to a tracelessness condition and to a divergenceless condition respectively. Obviously $e_{\mu\nu}$ is gauge invariant too:
\begin{equations}
& e_\mu{}^\mu = 0, \\
& \de^\mu\,e_{\mu\nu} = 0,\\
& \delta_\lambda\,e_{\mu\nu}=0.
\end{equations}
The definition of $e_{\mu\nu}$ can be extended to arbitrary spacetime dimensions:
\begin{equation}
e_{\mu_1\dots\mu_{d-2}\nu} := \tfrac{1}{3}\,\vepsilon_{\mu_1\dots\mu_{d-2}}{}^{\alpha\beta}\,F_{\nu\alpha\beta} = \vepsilon_{\mu_1\dots\mu_{d-2}}{}^{\alpha\beta}\,\de_\alpha\,h_{\beta\nu}\quad\text{in}\;d\;\text{dimensions},
\end{equation}
whose properties are
\begin{equations}
& e_{\mu_1\dots\mu_{d-3}\mu}{}^\mu = 0,\\
& \de^\mu\,e_{\mu\mu_1\dots\mu_{d-3}\nu} = 0,\\
& \delta_\lambda\,e_{\mu_1\dots\mu_{d-2}\nu} = 0.
\end{equations}
Notice that $e_{\mu_1\dots\mu_{d-2}\nu}$ selects the curl of $h_{\mu\nu}$:
\begin{equation}
e_{\mu_1\dots\mu_{d-2}\nu} = \tfrac{1}{2}\,\vepsilon_{\mu_1\dots\mu_{d-2}}{}^{\alpha\beta}\,\de_{[\alpha}\,h_{\beta]\nu}  = \tfrac{1}{2}\,\vepsilon_{\mu_1\dots\mu_{d-2}}{}^{\alpha\beta}\,f_{\alpha\beta\nu},
\end{equation}
where $f_{\mu\nu\vrho}$ denotes the curl
\begin{equation}
f_{\mu\nu\vrho} := \de_\mu\,h_{\nu\vrho} - \de_\nu\,h_{\mu\vrho}.
\end{equation}
The properties of $f_{\mu\nu\vrho}$ are
\begin{equations}
& f_{\mu\nu\vrho} + f_{\nu\vrho\mu} + f_{\vrho\mu\nu} = 0,\label{cyclicf}\\
& \de_\sigma\,f_{\mu\nu\vrho} + \de_\mu\,f_{\nu\sigma\vrho} + \de_\nu\,f_{\sigma\mu\vrho} = 0,\\
& \delta_\lambda\,f_{\mu\nu\vrho} = 0.
\end{equations}
The cyclicity and the Bianchi identity can be also written as
\begin{equations}
& \vepsilon^{\mu_1\dots\mu_{d-3}\mu\nu\vrho}\,f_{\mu\nu\vrho} = 0,\\
& \vepsilon^{\mu_1\dots\mu_{d-3}\sigma\mu\nu}\,\de_\sigma\,f_{\mu\nu\vrho} = 0.
\end{equations}
If $\de_\mu\,h_{\nu\vrho}$ is viewed as a connection, $f_{\mu\nu\vrho}$ is its torsion. This explains why $f_{\mu\nu\vrho}$ is gauge invariant.\footnote{Compare instead with the (linearized) Levi-Civita connection: $\Gamma_\mu{}^\vrho{}_\nu = \tfrac{1}{2}\,(\de_\mu\,h^\vrho{}_\nu + \de_\nu\,h^\vrho{}_\mu - \de^\vrho\,h_{\mu\nu})$, which, being symmetric, is torsionless.}

In the following, the relations between $F_{\mu\nu\vrho}$, $f_{\mu\nu\vrho}$ and $e_{\mu_1\dots\mu_{d-2}\nu}$ are summarized:
\begin{equations}
& e_{\mu_1\dots\mu_{d-2}\nu} = \tfrac{1}{2}\,\vepsilon_{\mu_1\,\dots\mu_{d-2}}{}^{\alpha\beta}\,f_{\alpha\beta\nu} = \tfrac{1}{3}\,\vepsilon_{\mu_1\dots\mu_{d-2}}{}^{\alpha\beta}\,F_{\nu\alpha\beta},\\
& f_{\alpha\beta\mu} = \tfrac{1}{3}\,F_{\mu[\alpha\beta]} = \tfrac{(-)^{\text{sgn}}}{(d-2)!}\,\vepsilon_{\alpha\beta}{}^{\mu_1\dots\mu_{d-2}}\,e_{\mu_1\dots\mu_{d-2}\mu},\\
& F_{\alpha\beta\mu} = - f_{\mu(\alpha\beta)} = -\tfrac{(-)^{\text{sgn}}}{(d-2)!}\,\vepsilon_{\mu\alpha}{}^{\mu_1\dots\mu_{d-2}}\,e_{\mu_1\dots\mu_{d-2}\beta} + (\alpha \leftrightarrow \beta),
\end{equations}
where $\text{sgn}$ is the number of negative eigenvalues of the flat metric. 

Let us conclude this section with a comment on the mass dimension of the fracton gauge field. When a field theory is placed on a curved background, the mass dimension or the rigid length scaling is equivalent to the grading induced by scaling the background metric tensor. Consider a free theory for a scalar field $\vphi$ in two dimensions. The action reads $\int\diff^2 x\,\de_\mu\,\vphi\,\de^\mu\,\vphi$. If the coordinates are rescaled according to $\delta\,x^\mu = - x^\mu$, so that $\diff x^\mu$ is rescaled by $-1$ and $\de_\mu$ by 1, then $\vphi$ must have vanishing scaling in order the action to be dimensionless. Equivalently, one can put the theory on a curved background, with metric $g_{\mu\nu}$, and rescale the metric. Indeed, $d s^2 = g_{\mu\nu}\,\diff x^\mu\,\diff x^\nu$, and if one rescales $\diff x^\mu$ by $-1$ \emph{or} $g_{\mu\nu}$ by $-2$, $\diff s^2$ is rescaled by $-2$ in both the cases. The action in the curved background reads $\int \diff^2 x\,\sqrt{|g|}\,g^{\mu\nu}\,\de_\mu\,\vphi\,\de_\nu\,\vphi$, and it is rigidly scale invariant, since the inverse metric $g^{\mu\nu}$ is scaled by $2$ and the square root of the determinant of the metric by $-2$ in two spacetime dimensions (in general, it is rescaled by minus the number of spacetime dimensions). 

The same phenomenon occurs in a theory of a free vector field $A_{\mu}$ with field strength $F_{\mu\nu} = \de_{[\mu}\,A_{\nu]}$ in four dimensions. The action, which reads $\int\diff^4 x\,F_{\mu\nu}\,F^{\mu\nu}$, is scale invariant if $A_\mu$ is scaled by 1. Equivalently, the action on a curved background, which reads $\int\diff^4 x\,\sqrt{|g|}\,g^{\mu\vrho}\,g^{\nu\sigma}\,F_{\mu\nu}\,F_{\vrho\sigma}$, is scale invariant upon rescaling the metric $g_{\mu\nu}$ as before. 

In this sense, we say that zero is the canonical mass dimension of the scalar field and one is the canonical mass dimension of the vector field
\begin{equation}
[\vphi] = 0, \quad [A_\mu] = 1,
\end{equation} 
since they are the mass dimensions which ensure the rigid scale invariance of the theory, in two and in four dimensions respectively, to be equivalent to the scale invariance on a curved background. In dimensions other than two or four, a dimensionful coupling constant has to be included in the action for maintaining the counting. This is the reason why the electric charge is dimensionless in four dimensions, but it is dimensionful in the other ones. 

This discussion allows one to assign a canonical dimension to the fracton gauge field. A quadratic action in the derivative of $h_{\mu\nu}$ on a curved background is scale invariant by rescaling the metric only in six dimensions, since in that case three inverse metrics are involved, compensating the scaling of the measure of integration: $\int\diff^6 x\,\sqrt{|g|}\,g^{\mu\vrho}g^{\nu\sigma}\,g^{\lambda\tau}\,f_{\mu\nu\lambda}\,f_{\vrho\sigma\tau}$. (Another contraction is possible in defining a gauge invariant action, but for the scaling argument this is not relevant.) The same result is obtained on a fixed flat background by rescaling $h_{\mu\nu}$ by $2$. So, we define the canonical mass dimension of the fracton gauge field as
\begin{equation}
[h_{\mu\nu}] = 2.
\end{equation} 
In spacetime dimensions other than six, a dimensionful coupling constant has to be included in the action. In general, the number of lower indices gives the canonical mass dimension of the field. In this sense, differential forms, obtained by contracting tensor components with differentials, have automatically vanishing mass dimension.

\section{BRST fracton anomalies}\label{Sec2}

The current $J^{\mu\nu}$ may be viewed as the Noether current associated to a global symmetry enjoyed by the matter theory. The conservation law \eqref{WeakConservation} is a consequence of the gauge invariance and of the Lorentz invariance of a minimal coupling between $h_{\mu\nu}$ and $J^{\mu\nu}$:
\begin{equation}\label{MinimalCoupling}
\int\diff^d x\,h_{\mu\nu}\,J^{\mu\nu},
\end{equation}
as in usual gauge theories. Indeed,
\begin{align}
\int\diff^d x\,h_{\mu\nu}\,J^{\mu\nu} &\rightarrow
\int\diff^d x\,\Lambda_\mu{}^\alpha\,\Lambda_\nu{}^\beta\,(h_{\alpha\beta} + \de_\alpha\,\de_\beta\,\lambda)\,\Lambda^\mu{}_\gamma\,\Lambda^\nu{}_\delta\,J^{\gamma\delta} =\nn\\
& = \int\diff^d x\, (h_{\mu\nu}\,J^{\mu\nu} + \de_\mu\,\de_\nu\,\lambda\,J^{\mu\nu}) = 
\int\diff^d x\, (h_{\mu\nu}\,J^{\mu\nu} + \lambda\,\de_\mu\,\de_\nu\,J^{\mu\nu}) = \nn\\
& = \int\diff^d x\, h_{\mu\nu}\,J^{\mu\nu} 
\Leftrightarrow \de_\mu\,\de_\nu\,J^{\mu\nu}  = 0,
\end{align}
where $\Lambda_\mu{}^\alpha$ is a Lorentz rotation.

The classical global symmetry is a symmetry of the quantized theory too, if there is no anomaly, that is, if the quantum expectation value of the current satisfies the conservation law. If this is not the case, there exists some breaking term $\mathcal{A}$, such that
\begin{equation}
\de_\mu\,\de_\nu\,\langle J^{\mu\nu}\rangle = \mathcal{A}.
\end{equation} 
$\int\diff^d x\,\lambda\,\mathcal{A}$ is an (integrated) anomaly by definition.  The theory is not prevented to be quantized, even if an anomaly is present, if the symmetry remains global, that is, if the gauge field is treated as a nonintegrated background field in the partition function:
\begin{equation}
e^{i\,\mathcal{W}[h]} = \int [\diff\vphi^i]\,e^{i\,\hat{S}[\vphi^i, h]},
\end{equation} 
where $\mathcal{W}[h]$ is the quantum effective action. Then, if the matter coupling is minimal, the quantum expectation value of the Noether current is equal to the first functional derivative of the quantum effective action with respect to the gauge field $h_{\mu\nu}$, evaluated at $h_{\mu\nu}= 0$:
\begin{equation}
\langle J^{\mu\nu}\rangle = {\frac{\delta\,\mathcal{W}[h]}{\delta\,h_{\mu\nu}}}{\bigg{|}_{h_{\mu\nu}=0}}.
\end{equation}

The possible consistent anomalies should depend only on the gauge sector, as a consequence of the Wess-Zumino consistency conditions. In the \textsc{brst} formulation \cite{Becchi:1974md, Becchi:1974xu, Becchi:1975nq, Tyutin:1975qk}, the consistency conditions are phrased in terms of a cohomology problem \cite{Stora:1976LM, Stora:1984, Zumino:1983ew, Zumino:1983rz}. 

Denote with $s$ the \textsc{brst} operator and promote $\lambda$ to an anticommuting ghost. $s$ should be a nilpotent, odd differential. The grade $g$ induced by $s$, called \emph{ghost number}, counts the number of ghosts in a differential $p$-form $\omega^{(p)}_g$. The action of $s$ on $\omega^{(p)}_g$ increases $g$ by one, as well as the de Rham differential increases the form degree $p$ by one. 

The nilpotent \textsc{brst} transformations in the fracton gauge sector read
\begin{equations}
s\,\lambda &= 0,\label{BRSTrule1}\\
s\,h_{\mu\nu} &= \de_\mu\,\de_\nu\,\lambda.\label{BRSTrule2}
\end{equations}
The anomaly $\int\diff^d x\,\lambda\,\mathcal{A}$ is an integrated $d$-form of ghost number one, which we denote with $\omega^{(d)}_1$. The consistency condition is
\begin{equation}
s\int \omega^{(d)}_1 = 0,
\end{equation}
or, for the nonintegrated anomaly,
\begin{equation}
s\,\omega^{(d)}_1 = -\,\diff \omega^{(d-1)}_2,
\end{equation}
for some $\omega^{(d-1)}_2$. The $s$ nilpotency, the anticommutativity between $s$ and $\diff$ and the local triviality of the de Rham cohomology, implies a descent of equations:
\begin{equations} 
s\,\omega_1^{(d)} &= - \,\diff \omega_2^{(d-1)},\label{descent1}\\
s\,\omega_2^{(d-1)} &= - \,\diff \omega_3^{(d-2)},\\
&\vdots\\
s\,\omega_d^{(1)} &= - \,\diff \omega_{d+1}^{(0)},\\
s\,\omega_{d+1}^{(0)} &= 0,\label{descent2}
\end{equations}
where $\omega_{d+1-p}^{(p)}$ is a $p$-form of ghost number $d+1-p$. The aim is to compute the nontrivial anomalies $\omega_1^{(d)}$, by starting from a nontrivial $\omega_{d+1}^{(0)}$, which is an element of the $s$ cohomology on zero-forms of ghost number $d+1$, and rising the descent.

\subsection{BRST fracton cohomology}\label{BRSTcoho}

The first step in solving the descent is to compute the $s$ cohomology as the ghost number varies, that is, one has to solve the equations
\begin{equation}
s\,\omega_{d+1-p}^{(p)} = 0, \quad p=0,\dots, d.
\end{equation}

Consider the \textsc{brst} transformations for the multiple derivatives of $h_{\mu\nu}$:
\begin{equations}
s\,\de_{\alpha_1}\dots\de_{\alpha_p}\,h_{\mu\nu} &= \de_{\alpha_1}\dots\de_{\alpha_p}\,\de_\mu\,\de_\nu\,\lambda,\\
s\,\de_{\alpha_1}\dots\de_{\alpha_p}\,\de_\mu\,\de_\nu\,\lambda &= 0.
\end{equations}
Because of the symmetries of the right-hand side, the first one splits into,\footnote{Here and in the following the bracket among the indices for symmetrization $(\dots)$ or antisymmetrization $[\dots]$ do \emph{not} include any numerical factor.}
\begin{equations}
s\,[\tfrac{1}{(p+2)!}\,\de_{(\alpha_1}\dots\de_{\alpha_p}\,h_{\mu\nu)}] &= \de_{\alpha_1}\dots\de_{\alpha_p}\,\de_\mu\,\de_\nu\,\lambda,\\
s\,\hat{h}_{\mu\nu,\alpha_1\dots\alpha_p} &= 0,
\end{equations}
where we defined
\begin{equation}
\hat{h}_{\mu\nu,\alpha_1\dots\alpha_p} := \de_{\alpha_1}\dots\de_{\alpha_p}\,h_{\mu\nu}-\tfrac{1}{(p+2)!}\,\de_{(\alpha_1}\,\dots\de_{\alpha_p}\,h_{\mu\nu)}.
\end{equation}
Therefore $\{\tfrac{1}{(p+2)!}\,\de_{(\alpha_1}\,\dots\de_{\alpha_p}\,h_{\mu\nu)},\de_{\alpha_1}\dots\de_{\alpha_p}\,\de_\mu\,\de_\nu\,\lambda\}$ is a doublet for each $p$, so it is pulled out from the $s$ cohomology by filtering arguments \cite{Brandt:1989rd, Piguet:1995er}, and the $s$ cohomology is generated by $\{\lambda, \de_\mu\,\lambda, \hat{h}_{\mu\nu,\alpha}, \hat{h}_{\mu\nu,\alpha\beta}, \dots \}$. Actually, $\hat{h}_{\mu\nu,\alpha_1\dots\alpha_p}$ can be replaced by the derivatives of $f_{\mu\nu\alpha}$. Indeed, $\hat{h}_{\mu\nu,\alpha_1\dots\alpha_p}$ can be expressed in terms of the derivatives of $f_{\mu\nu\alpha}$, for each $p$, as we prove at the end of this section. So, the $s$ cohomology on local polynomials in the fields of the theory can be chosen to be generated by 
\begin{equation}
\{\lambda, \lambda_\mu, f_{\mu\nu\alpha}, \de_\beta\,f_{\mu\nu\alpha}, \de_\beta\,\de_\gamma\,f_{\mu\nu\alpha},\dots \},
\end{equation}
where we defined
\begin{equation}
\lambda_\mu := \de_\mu\,\lambda.
\end{equation}
We denote with ${\omega^\natural}^{(p)}_g$ a $p$-forms of ghost number $g$ in the $s$ cohomology:
\begin{equation}
s\,{\omega^{\natural}}^{(p)}_g = 0, \quad
{\omega^{\natural}}^{(p)}_g \neq s\,\psi^{(p)}_{g-1},
\end{equation}
for any $p$-form of ghost number $g-1$ $\psi^{(p)}_{g-1}$. In order to find the most general ${\omega^\natural}^{(p)}_{d+1-p}$, one has to take into account the mass dimensions of the generators:
\begin{equation}
[\lambda]= 0,\quad
[\lambda_\mu]= 1,\quad
[f_{\mu\nu\vrho}]= 3,\quad
[\de_\sigma\,f_{\mu\nu\vrho}]= 4, \;\;\dots
\end{equation}
The mass dimension of the ghost is fixed by the\textsc{brst} transformations, since $s$ preserves the mass dimension. Moreover, notice that 
\begin{equation}\label{anticommutativity_lambda}
\lambda^2 = 0, \quad \lambda_\mu\,\lambda^\mu  = 0.
\end{equation}
because of anticommutativity. Furthermore,
\begin{equation}\label{anticommutativity_lambda2}
\lambda_{\mu_1}\,\lambda_{\mu_2}\dots\,\lambda_{\mu_{d+1}} = 0 \;\;\text{in}\;d\,\text{dimensions},
\end{equation}
because in $d$ dimensions we have at most $d$ possible $\lambda_\mu$'s, so that one of them necessarily repeats in a product of $d+1$ ones.

Let us conclude this section by proving that each $\hat{h}_{\mu\nu,\alpha\alpha_1\dots\alpha_p}$ can be written in terms of the derivatives of $f_{\mu\nu\alpha}$, producing an explicit formula (although this is not strictly necessary for our purposes). Observe that
\begin{align}
\vepsilon_{\mu_1\dots\mu_{d-2}}{}^{\alpha\mu}\,\hat{h}_{\mu\nu,\alpha\alpha_1\dots\alpha_p} &= 
\vepsilon_{\mu_1\dots\mu_{d-2}}{}^{\alpha\mu}\,\de_{\alpha_1}\dots\de_{\alpha_p}\,\de_\alpha\,h_{\mu\nu} = \nn\\
& = \tfrac{1}{2}\,\vepsilon_{\mu_1\dots\mu_{d-2}}{}^{\alpha\mu}\,\de_{\alpha_1}\dots\de_{\alpha_p}\,f_{\alpha\mu\nu},
\end{align}
Therefore, $\hat{h}_{\mu\nu,\alpha\alpha_1\dots\alpha_p}$ is a linear combination of $\de_{\alpha_1}\dots\de_{\alpha_p}\,f_{\alpha\mu\nu}$, up to permutations of the indices. The ansatz should be 
\begin{equation}
\hat{h}_{\mu\nu,\alpha\alpha_1\dots\alpha_p} = 
A_{1,p}\,\de_{(\alpha_1}\dots\de_{\alpha_p}\,f_{\alpha)(\mu\nu)} + A_{2,p}\,\de_{(\alpha_1}\dots\de_{\alpha_{p-1}|}\,\de_{(\mu}\,f_{\nu)|\alpha_p\alpha)},
\end{equation}
since in this way the symmetries of $\hat{h}_{\mu\nu,\alpha\alpha_1\dots\alpha_p}$ are satisfied in the right-hand side. The coefficients $A_{1,p}$ and $A_{2,p}$ are fixed by expanding the left-hand side and the right-hand side in terms of the derivatives of $h_{\mu\nu}$. Namely, for the left-hand side
\begin{align}
\hat{h}_{\mu\nu,\alpha\alpha_1\dots\alpha_p} &= 
(1-\tfrac{1}{N_p})\,\de_{\alpha_1}\dots\de_{\alpha_p}\,\de_\alpha\,h_{\mu\nu} \,+\nn\\
& - \tfrac{1}{N_p}[\de_{(\alpha_1}\dots\de_{\alpha_p|}\,\de_{(\mu}\,h_{\nu)|\alpha)} + \de_{\mu}\de_{\nu}\,\de_{(\alpha_1}\dots\de_{\alpha_{p-1}}\,h_{\alpha_p\alpha)}],
\end{align}
where $N_p = (p+2)(p+3)/2$. Instead, expanding the right-hand side, one sees that the coefficient of the term $\de_{\alpha_1}\dots\de_{\alpha_p}\,\de_\alpha\,h_{\mu\nu}$ is $(p+1)!\,2!\,A_{1,p}$, and the coefficient of the terms $\de_{\mu}\,\de_{\nu}\,\de_{\cdot}\cdots\de_{\cdot}\,h_{\cdot\cdot}$ is $(p-1)!\,2!\,2!\,A_{2,p}$. Therefore, we conclude that 
\begin{equation}
A_{1,p} = \tfrac{N_p-1}{(p+1)!\,2\,N_p}, \quad
A_{2,p} = -\tfrac{1}{(p+1)!\,4\,N_p}, \quad
N_p = \tfrac{(p+2)(p+3)}{2} \quad \text{with}\;\;p=1,2,\dots
\end{equation}
In particular, the first few cases are 
\begin{equations}
\hat{h}_{\mu\nu,\alpha\beta}  &= \tfrac{5}{24}\,\de_{(\beta}\,f_{\alpha)(\mu\nu)} -\tfrac{1}{24}\,\de_{(\mu}\,f_{\nu)(\beta\alpha)},\\
\hat{h}_{\mu\nu,\alpha\beta\gamma} &= \tfrac{3}{40}\,\de_{(\beta}\,\de_\gamma\,f_{\alpha)(\mu\nu)} -\tfrac{1}{40}\,\de_{(\beta|}\de_{(\mu}\,f_{\nu)|\gamma\alpha)},\\
\hat{h}_{\mu\nu,\alpha\beta\gamma\delta} &= \tfrac{7}{360}\,\de_{(\beta}\,\de_\gamma\,\de_\delta\,f_{\alpha)(\mu\nu)} -\tfrac{1}{120}\,\de_{(\beta}\de_{\gamma|}\de_{(\mu}\,f_{\nu)|\delta\alpha)},\\
\hat{h}_{\mu\nu,\alpha\beta\gamma\delta\vepsilon} &=
\tfrac{1}{252}\,\de_{(\beta}\,\de_\gamma\,\de_\delta\de_{\vepsilon}\,f_{\alpha)(\mu\nu)} -\tfrac{1}{504}\,\de_{(\beta}\de_{\gamma}\,\de_{\delta|}\,\de_{(\mu}\,f_{\nu)|\vepsilon\alpha)}.
\end{equations}
The peculiar case $p=0$ follows by explicit computation:
\begin{align}
\hat{h}_{\mu\nu,\alpha} &= \de_\alpha\,h_{\mu\nu}-\tfrac{1}{6}\,\de_{(\alpha}\,h_{\mu\nu)} = 
(1-\tfrac{1}{3})\,\de_\alpha\,h_{\mu\nu} - \tfrac{1}{3}\,\de_{(\mu}\,h_{\nu)\alpha} = \nn\\
&= -\tfrac{1}{3}\,(\de_{(\mu}\,h_{\nu)\alpha} - 2\,\de_{\alpha}\,h_{\mu\nu}) = -\tfrac{1}{3}\,F_{\mu\nu\alpha} = \tfrac{1}{3}\,f_{\alpha(\mu\nu)}.
\end{align}

\subsection{Construction of the cohomology classes}\label{cohoClasses}

In solving the $d$-dimensional descent we are interested in the $s$ cohomology on the $p$-forms of ghost number $d+1-p$, where $0 \leqslant p \leqslant d$. Let us denote an arbitrary cohomology class in that sector with ${\omega^{\natural}}^{(p)}_{d+1-p}$. The most general expression reads
\begin{align}
{\omega^{\natural}}^{(p)}_{d+1-p} &= A^{\mu_1\dots\mu_{d-p}}{}_{\alpha_1\dots\alpha_p}\,\lambda\,\de_{\mu_1}\,\lambda\dots\de_{\mu_{d-p}}\,\lambda\,\diff x^{\alpha_1} \dots \diff x^{\alpha_p} \,+\nn\\
& + B^{\mu_1\dots\mu_{d+1-p}}{}_{\alpha_1\dots\alpha_p}\,\de_{\mu_1}\,\lambda\dots\de_{\mu_{d+1-p}}\,\lambda\,\diff x^{\alpha_1} \dots \diff x^{\alpha_p}, 
\end{align}
where $A^{\mu_1\dots\mu_{d-p}}{}_{\alpha_1\dots\alpha_p}$ and $B^{\mu_1\dots\mu_{d+1-p}}{}_{\alpha_1\dots\alpha_p}$ are totally antisymmetric in the $\mu$'s and $\alpha$'s, having mass dimension $2\,p-d$ and $2\,p-1-d$ respectively. Since we can only use $\eta_{\mu\nu}$, $\vepsilon_{\mu_1\dots\mu_d}$ and $f_{\alpha\beta\mu}$ and its derivatives to construct them, and $f_{\alpha\beta\mu}$ has mass dimension three, $A^{\mu_1\dots\mu_{d-p}}{}_{\alpha_1\dots\alpha_p}$ and $B^{\mu_1\dots\mu_{d+1-p}}{}_{\alpha_1\dots\alpha_p}$ have vanishing mass dimension or at least mass dimension three. $A^{\mu_1\dots\mu_{d-p}}{}_{\alpha_1\dots\alpha_p}$ has vanishing mass dimension if and only if the number of spacetime dimensions is even $d = 2\,n$, and $p=n$. In this case, $A^{\mu_1\dots\mu_n}{}_{\alpha_1\dots\alpha_n}$ can be chosen proportional to $\delta^{\mu_1}_{[\alpha_1}\dots \delta^{\mu_n}_{\alpha_n]}$ or $\vepsilon^{\mu_1\dots\mu_n}{}_{\alpha_1\dots\alpha_n}$. The two possibilities cannot mix together, since they have opposite Lorentz parity, which is preserved by the \textsc{brst} operator. The corresponding $B^{\mu_1\dots\mu_{n+1}}{}_{\alpha_1\dots\alpha_n}$ vanishes, since it has mass dimension $-1$. Similarly, $B^{\mu_1\dots\mu_{d+1-p}}{}_{\alpha_1\dots\alpha_p}$ has vanishing mass dimension if and only if the number of spacetime dimensions is odd $d = 2\,n - 1$, and $p=n$. In this case $B^{\mu_1\dots\mu_n}{}_{\alpha_1\dots\alpha_n}$ can be only chosen proportional to $\delta^{\mu_1}_{[\alpha_1}\dots \delta^{\mu_n}_{\alpha_n]}$, since the Levi-Civita tensor has an odd number of indices. The corresponding $A^{\mu_1\dots\mu_{n-1}}{}_{\alpha_1\dots\alpha_n}$ vanishes, having mass dimension 1. Summarizing, we have two classes on $n$-forms with opposite parity in even spacetime dimensions and a single class on $n$-forms in odd spacetime dimensions:
\begin{equations}
& {\omega^{\natural}}^{(n)}_{n+1} = \lambda\,\diff\lambda\dots\diff\lambda, \quad
{\tilde{\omega}}^{\natural(n)}_{n+1} = \lambda\,{\star(\diff\lambda\dots\diff\lambda)}, \quad \text{if}\; d = 2\,n, \label{EvenCohoClass}\\
& {\omega^{\natural}}^{(n)}_{n} = \diff\lambda\dots\diff\lambda, \quad \text{if}\; d = 2\,n-1,\label{OddCohoClass}
\end{equations}
where $\star$ denotes the Hodge dual and
\begin{equation}
\diff \lambda = \diff x^\mu\,\de_\mu\,\lambda\quad {\star\diff\lambda} = \de_\mu\,\lambda\,\vepsilon^\mu{}_{\alpha_1\dots\alpha_{d-1}}\,\diff x^{\alpha_1}\dots\diff x^{\alpha_{d-1}},
\end{equation}
(notice that there is an ambiguity in the sign, fixed by convention, since $\de_\mu\,\lambda$ and $\diff x^\alpha$ anticommute).

Otherwise, $A^{\mu_1\dots\mu_{d-p}}{}_{\alpha_1\dots\alpha_p}$ and $B^{\mu_1\dots\mu_{d-1-p}}{}_{\alpha_1\dots\alpha_p}$ could not vanish if $ n + 2 \leqslant p \leqslant d$.  Writing $p = 3\,\lfloor\frac{p}{3}\rfloor + r_p$, where $r_p$ is the integer remainder in the division by three, $A^{\mu_1\dots\mu_{d-p}}{}_{\alpha_1\dots\alpha_p}$ should contain $\lfloor \frac{2p-d}{3}\rfloor$ $f$'s and $r_{2p-d}$ derivatives; similarly, $B^{\mu_1\dots\mu_{d-1-p}}{}_{\alpha_1\dots\alpha_p}$ should contain $\lfloor \frac{2p-1-d}{3}\rfloor$ $f$'s and $r_{2p-1-d}$ derivatives. Let us consider the first examples:
\begin{itemize}
\item[--] If $d=2$, the cohomology on the zero-forms of ghost number three and the cohomology on the two-forms of ghost number one are trivial. The cohomology on the one-forms of ghost number two has two possible classes, with opposite parity, $\lambda\,\diff\lambda$ or $\lambda\,{\star\diff\lambda}$.
\item[--] If $d=3$, the cohomology on the zero-forms of ghost number four and the cohomology on the one-forms of ghost number three are trivial; the cohomology on the two-forms of ghost number two has a unique class $\diff\lambda\,\diff \lambda$. In the cohomology on the three-forms of ghost number one, the dimensional analysis selects the structures $f\,\lambda\,(\diff x)^3$ or $f\,\lambda\,\vepsilon\,(\diff x)^3$, but both vanish, since $f_{[\alpha\beta\mu]}=0$ in the first case, and since an odd number of indices to be contracted is involved in the second case. Therefore, the cohomology on three-forms of ghost number one is trivial. 
\item[--] If $d=4$, the cohomology on the zero-, one- and three-forms of ghost number five, four, and two, respectively, are trivial; the cohomology on the two-forms of ghost number three has two classes with opposite parity, $\lambda\,\diff\lambda\,\diff\lambda$ and $\lambda\,{\star(\diff\lambda\,\diff\lambda)}$. The cohomology on the four-forms of ghost number one is given by the possible contractions of $\de\,f\,\,\lambda\,\vepsilon\,(\diff x)^4$ and $f\,\de\,\lambda\,\vepsilon\,(\diff x)^4$, where the Levi-Civita tensor has to be included, since $f_{[\alpha\beta\mu]}=0$.
\item[--] If $d=5$, the cohomology on the zero-, one- and two-forms of ghost number six, five and four are trivial. The cohomology on the three-forms of ghost number three is given by $\diff\lambda\,\diff\lambda\,\diff \lambda$. The cohomology on the four-forms of ghost number two has a single structure $f\,\lambda\,\de\lambda\,(\diff x)^4$ (no opposite parity structure is allowed since the number of indices to be saturated is odd), but it vanishes, since $f_{[\alpha\beta\mu]}=0$. The cohomology on the five-forms of ghost number one, whose structure is
$\de\,\de\,f\,\lambda\,(\diff x)^5 + \de\,f\,\de\,\lambda\,(\diff x)^5$, is trivial for the same reason.
\item[--] If $d=6$, the cohomology on the zero-, one-, two-, and four-forms of ghost number seven, six, five and three respectively are trivial. The cohomology on the three-forms of ghost number four has two classes with opposite parity, $\lambda\,\diff\lambda\,\diff\lambda\,\diff\lambda$ and $\lambda\,{\star(\diff\lambda\,\diff\lambda\,\diff\lambda)}$. The most general class on the five-forms of ghost number two has the structure $\de\,f\,\lambda\,\de\,\lambda\,\vepsilon\,(\diff x)^5 + f\,(\de\,\lambda)^2\,\vepsilon\,(\diff x)^5$, where the omitted indices are saturated in all the possible ways. Similarly, the most general class on the six-forms of ghost number one has three possible structures $f\,f\,\lambda\,\vepsilon\,(\diff x)^6 + \de^3\,f\,\lambda\,\vepsilon\,(\diff x)^6 + \de^2\,f\,\de\,\lambda\,\vepsilon\,(\diff x)^6$.  
\end{itemize}
As a general lesson, we see that, in odd spacetime dimensions, the unique nontrivial cohomology class is ${\omega^{\natural}}^{(n)}_n$ in \eqref{OddCohoClass}. Notice that it is a total derivative, being equal to $\diff(\lambda\,(\diff\lambda)^{n-1})$. Therefore, the equation $s\,\omega^{(n+1)}_{n-1} = -\diff {\omega^{\natural}}^{(n)}_n = 0$ is uniquely solved by the cohomology class ${\omega^{\natural}}^{(n+1)}_{n-1}$. But this sector is empty. So, there is no nontrivial solution of the descent in odd spacetime dimensions. This means that \emph{there is no anomaly in covariant fracton theories in odd spacetime dimensions}, as in all the gauge theories in odd spacetime dimensions.

In even spacetime dimensions, the cohomology starts to be nontrivial on the $n$-forms of ghost number $n+1$, as in \eqref{EvenCohoClass}. The first class $\lambda\,(\diff\lambda)^n$ does not provide a solution of the descent. Indeed, its external differential $(\diff\lambda)^{n+1}$ sits in the cohomology on $n+1$ forms of the same ghost number, so that it cannot be written as an $s$ variation. As we show in the subsequent sections, the opposite parity class $\lambda\,{\star(\diff\lambda)^n}$ provides a solution of the descent only in the two-dimensional case. Therefore, if $d\geqslant 4$, the descent starts with $s\,\omega^{(n+2)}_{n-1} = 0$, $s\,\omega^{(n+3)}_{n-2} = -\,\diff \omega^{(n+2)}_{n-1}$. Indeed, the cohomology on the $n+1$ forms of ghost number $n$ is trivial. 

\subsection{Polyform formalism}\label{polyforms}

It is useful for the following to introduce a formalism which allows one to replace the whole descent
\eqref{descent1}--\eqref{descent2} by a single cohomology equation. In this way, it is simple to obtain results valid for every spacetime dimensions. This formalism, first introduced by Stora and Zumino \cite{Stora:1976LM, Stora:1984, Zumino:1983ew, Manes:1985df}, is based upon defining polyforms, whose total degree is given by the sum of the form degree and of the ghost number \cite{Zumino:1983rz, Imbimbo:2023sph}. The equation replacing the descent, called \emph{Stora-Zumino equation}, reads
\begin{equation}\label{Stora-Zumino}
\delta\,\omega_{d+1} = 0,
\end{equation}
where $\omega_{d+1}$ is the polyform of total degree $d+1$ encoding all the differential forms involved in the descent:
\begin{equation}
\omega_{d+1} := \omega^{(d)}_1 + \omega^{(d-1)}_2 + \dots + \omega^{(0)}_{d+1}
\end{equation}
and $\delta$ is the nilpotent, odd differential \cite{Thierry-Mieg:1979fvq}
\begin{equation}
\delta := \diff + s.
\end{equation}
The nilpotency holds, since both $s$ and $\diff$ are nilpotent and they anticommute. The descent is obtained starting from the Stora-Zumino equation by filtering as the ghost number varies, considering that $s$ increases the ghost number by one and $\diff$ does not change it. The equivalence of the Stora-Zumino equation and the descent means that the $s$ cohomology modulo $\diff$ is equivalent to the $\delta$ cohomology. 

Let us define the following differential forms:
\begin{equations}
& h^{(1)}_\mu := h_{\alpha\mu}\,\diff x^\alpha, \\
& f^{(2)}_\mu := \tfrac{1}{2}\,f_{\alpha\beta\mu}\,\diff x^\alpha\,\diff x^\beta.
\end{equations}
$h^{(1)}_\mu$ and $f^{(2)}_\mu$ are, respectively a  one-form and a two-form of ghost number zero, taking values in the cotangent bundle of the spacetime. Their mass dimensions are
\begin{equation}
[h^{(1)}_\mu] = [f^{(2)}_\mu] = 1.
\end{equation}
$f^{(2)}_\mu$ is the external differential of $h^{(1)}_\mu$ by definition: 
\begin{equation}
f^{(2)}_\mu = \diff h^{(1)}_\mu.
\end{equation}
making manifest the fact that $f^{(2)}_\mu$ is the analogue of the torsion in linearized gravity.\footnote{Compare with the definition of torsion out of the vielbein in the Cartan formalism: $T^a = \text{D}\,e^a = \diff e^a + \omega^{ab}\,e_b $, $e^a$ being the 1-form vielbein and $\omega^{ab}$ the one-form spin connection. In the linearized (Abelian) theory, the second term drops out.} The differential Bianchi identity is a consequence of the nilpotency of the de Rham differential:
\begin{equation}
\diff f^{(2)}_\mu = 0.
\end{equation}
Notice that the following relations hold
\begin{equations}
\diff x^\mu\,h^{(1)}_\mu = 0, \label{diffh}\\ 
\diff x^\mu\,f^{(2)}_\mu = 0, \label{difff}
\end{equations}
since $h_{\mu\alpha}$ is symmetric and the totally antisymmetric part of $f_{\alpha\beta\mu}$ vanishes, thanks to \eqref{cyclicf}. The \textsc{brst} transformations can be rewritten as 
\begin{equations}
& s\,h^{(1)}_\mu = -\,\diff\lambda_\mu,\\ 
& s\,\lambda = 0,\\
& s\,f^{(2)}_\mu = 0,
\end{equations}
where $\lambda_\mu = \de_\mu\,\lambda$, as in the previous sections, and the minus sign in the $h^{(1)}_\mu$ transformation is due to the fact that $\diff x^\alpha$ anticommutes with $\lambda_\mu$. 

The following polyform of total degree one can be defined
\begin{equation}
H_\mu := h^{(1)}_\mu + \lambda_\mu.
\end{equation}
Its $\delta$ variation is
\begin{align}
\delta\,H_\mu &= (\diff + s)\,(h^{(1)}_\mu + \lambda_\mu) = \nn\\
& = \diff h^{(1)}_\mu + (s\,h^{(1)}_\mu + \diff\,\lambda_\mu) + s\,\lambda_\mu,
\end{align}
which means that the \textsc{brst} transformations and the definition of $f^{(2)}_\mu$ are equivalent to\footnote{This can be viewed as a \emph{horizontality condition}, just as in the Yang-Mills case, since $\delta\,H_\mu$ is the natural extension of $\diff h^{(1)}_\mu$ to polyforms \cite{Thierry-Mieg:1979fvq}. A polyform is said to be \emph{horizontal} if its components of ghost number higher than zero vanish.}
\begin{equation}
\delta\,H_\mu = f^{(2)}_\mu.
\end{equation}
The polyform extension of the Bianchi identity holds by means of the $\delta$ nilpotency:
\begin{equation}
\delta\,f^{(2)}_\mu = 0,
\end{equation}
which encodes the $s$ invariance of $f^{(2)}_\mu$ and the Bianchi identity.

The $\delta$ variation of the ghost $\lambda$, which can be viewed as a single component polyform of total degree one, is simply
\begin{equation}
\delta\,\lambda = \diff\lambda.
\end{equation}
In the following table, the total degree $\text{deg}$ and the mass dimension $\text{dim}$ of fields and operators are summarized for convenience:
\begin{table}[h]
\begin{tabular}{llllllll}
& $H_\mu$ & $f^{(2)}_\mu $ & $\lambda $ & $\diff\lambda $ & $\;\diff x^\mu $ & $\delta $ & $\de_\mu $ \\
deg & $1$ & $2$ & $1$ & $\;2$ & $\;\;\,1$ & $1$ & $0$ \\
dim  & $1$ & $1$ & $0$ & $\;0$& $-1$ & $0$ &$1$
\end{tabular}
\end{table}

In conclusion, the action of $\delta$ on the space $\{ H_\mu, f^{(2)}_\mu, \lambda, \diff\lambda \}$ consists into two doublets:
\begin{equation}
\delta\,H_\mu = f^{(2)}_\mu, \quad
\delta\,f^{(2)}_\mu = 0, \quad
\delta\,\lambda = \diff\lambda, \quad
\delta\,\diff\lambda= 0.
\end{equation}
So, the $\delta$ cohomology seems to be trivial: Each $\delta$ cocycle $P_n$ of total degree $n$, such that $\delta\,P_n = 0$, is a $\delta$ coboundary, that is, $P = \delta\,Q_{n-1}$, for some polyform $Q_{n-1}$ of total degree $n-1$. Nevertheless, if we specify the number of spacetime dimensions $d$, there are some vanishing polyforms. So we are actually working on a quotient space, on which the cohomology is not guaranteed to be trivial.\footnote{The author thanks Prof. Camillo Imbimbo for having pointed this out.} Moreover, passing in components, the two doublets are not independent.

In general, a polyform of total degree $>d$ is not necessarily vanishing in $d$ spacetime dimensions, since it may contain components of nonvanishing ghost number, which are eventually differential forms of degree $<d$. In our case, we can produce vanishing polyforms in $d$ spacetime dimensions by arranging $H_\mu$, $f^{(2)}_\mu$ and $\diff\lambda$ in a polyform whose minimum form degree is at least $d+1$. For example, each polyform involving at least $d + 1$ $\diff x^\mu\,H_\mu$ vanishes in $d$ dimensions, 
\begin{equation}
(\diff x^\mu\,H_\mu)^{d+1} = 0\quad\text{in}\;d\;\text{dimensions,}
\end{equation}
since the following identity holds
\begin{equation}\label{dlambda}
\diff x^\mu\,H_\mu = \diff \lambda,
\end{equation}
(this is a direct consequence of the symmetry of $h_{\alpha\mu}$), so that $\diff x^\mu\,H_\mu$ is actually a one-form of ghost number one. Moreover, observe that 
\begin{equation}
H_{\mu_1}\,\dots H_{\mu_{d+1}} = 0 \quad\text{in}\;d\;\text{dimensions.}
\end{equation}
for the same reason why the the dimensional-dependent identity \eqref{anticommutativity_lambda2} holds. The identity is true for any possible arrangements of the indices. 

\subsection{Some anomalies using polyforms}\label{SomeSolutions}

Let us find possible solutions of the Stora-Zumino equation \eqref{Stora-Zumino}. Consider $d=2$. It is quite natural to consider a polyform enhancement of the cohomology class $\lambda\,{\star\diff \lambda}$. We may pick up $\lambda\,H_\mu\,{\star\diff x^\mu}$, which is a polyform of total degree three and vanishing mass dimension, where ${\star\diff x^\mu}=\vepsilon^\mu{}_\nu\,\diff x^\nu$, whose component of ghost number one is precisely $\lambda\,{\star\diff \lambda}$. Observe that
\begin{equation}
\delta\,(H_\mu\,{\star\diff x^\mu}) = f_\mu^{(2)}\,{\star\diff x^\mu} = 0,
\end{equation}
since it is a three-form in two dimensions. Therefore,\footnote{One can also perform the computation in components: $\delta\,(\lambda\,H_\mu\,{\star\diff x^\mu}) = 
\diff\lambda\,H_\mu\,{\star\diff x^\mu} = 
\diff\lambda\,(h_\mu^{(1)} + \de_\mu\,\lambda)\,{\star\diff x^\mu}$. The first term vanishes, since it is three-form in two dimensions; the second term, which is equal to $\diff\lambda\,{\star\diff \lambda}$, vanishes too, since $\lambda$ is anticommuting.}
\begin{equation}
\delta\,(\lambda\,H_\mu\,{\star\diff x^\mu}) = 
H_\nu\,\diff x^\nu\,H_\mu\,{\star\diff x^\mu} = H_\mu\,H_\nu\,\vepsilon^\mu{}_\vrho\,\diff x^\nu\,\diff x^\vrho,
\end{equation}
where we used the property \eqref{dlambda}. But the previous expression vanishes, since it is dual to $H_\mu\,H^\mu = 0$ ($H_\mu$ is anticommuting): 
\begin{equation}
\delta\,(\lambda\,H_\mu\,{\star\diff x^\mu}) = 0.
\end{equation}
Thus, the component of ghost number one of $\lambda\,H_\mu\,{\star\diff x^\mu}$, which reads
\begin{equation}\label{Anomaly2d}
\omega^{(2)}_1 = \lambda\,\vepsilon^{\mu\nu}\,h^{(1)}_\mu\,\diff x_\nu,
\end{equation}
is an anomaly. Taking the dual, it is proportional to $\lambda\,h_\mu{}^\mu\,\diff^2 x$. Notice that the general discussion on the $s$ cohomology (the sector on the zero-forms of ghost number three and that on the two-forms of ghost number one in two dimensions are trivial, and the cohomology class $\lambda\,\diff\lambda$ does not provide a solution of the descent) shows that the previous anomaly is the \emph{unique} nontrivial  consistent anomaly in two dimensions.

In order to extend the previous result to any even spacetime dimensions $d= 2\,n$, one may consider $\lambda\,H_{\mu_1}\dots H_{\mu_n}\,{\star(\diff x^{\mu_1}\dots \diff x^{\mu_n})}$, since its component of maximum ghost number is $\lambda\,{\star(\diff\lambda)^n}$. But it is easy to see that it fails to be $\delta$ invariant. This allows one to conclude that $\lambda\,{\star(\diff\lambda)^n}$ alone does not bring to a solution of the descent, if $d \geqslant 4$ (See also Section \ref{Weyl}).

Nevertheless, if $\lambda\,H_{\mu}$ is replaced by $f^{(2)}_\mu$, preserving the total degree and the mass dimension, one gets a $\delta$ cocycle for any $n \geqslant 2$. Indeed,
\begin{align}
& \delta\,(f^{(2)}_{\mu_1}\,H_{\mu_2}\dots H_{\mu_n}\,{\star(\diff x^{\mu_1}\,\diff x^{\mu_2}\dots\diff x^{\mu_n})}) = \nn\\
= &\,(n-1)\,f^{(2)}_{\mu_1}\,f^{(2)}_{\mu_2}\,H_{\mu_3}\dots H_{\mu_n}\,{\star(\diff x^{\mu_1}\,\diff x^{\mu_2}\,\diff x^{\mu_3}\dots\diff x^{\mu_n})} = 0,
\end{align}
which vanishes since $f^{(2)}_{\mu_1}\,f^{(2)}_{\mu_2}$ is symmetric and $\vepsilon^{\mu_1\mu_2\dots}$ antisymmetric. This is a one-line proof that the component of ghost number one
\begin{align}\label{Anomaly2nd}
\omega_1^{(2n)} &= (n-1)\,f^{(2)}_{\mu_1}\,h_{\mu_2}^{(1)}\dots h_{\mu_{n-1}}^{(1)}\,\de_{\mu_n}\,\lambda\,{\star(\diff x^{\mu_1}\,\diff x^{\mu_2}\dots\diff x^{\mu_{n-1}}\diff x^{\mu_n})},
\end{align}
that is
\begin{align}
\omega_1^{(2n)} &= \tfrac{1}{n(n-2)!}\,f^{(2)}_{\mu_1}\,h_{\mu_2}^{(1)}\dots h_{\mu_{n-1}}^{(1)}\,\de_{\mu_n}\,\lambda\,\times\nn\\
& \times\vepsilon^{\mu_1\mu_2\dots\mu_{n-1}\mu_n}{}_{\alpha_1\alpha_2\dots\alpha_{n-1}\alpha_n}\,\diff x^{\alpha_1}\,\diff x^{\alpha_2}\dots\diff x^{\alpha_{n-1}}\diff x^{\alpha_n},
\end{align}
is an anomaly in $2\,n \geqslant 4$ spacetime dimensions.

Many other anomalies are, in principle, allowed if we consider also the Hodge dual of $f_\mu^{(2)}$, or contractions involving the form indices of $f^{(2)}_\mu$. These contractions are unavoidably dimensional dependent. 

Nevertheless, we said before that the anomaly \eqref{Anomaly2d} is the most general in two dimensions, and the same claim is also valid in four dimensions. Let us see why. As it follows from the general discussion of the fracton \textsc{brst} cohomology, in four dimensions the problem consists simply in writing down the most general class in the \textsc{brst} cohomology on the four-forms of ghost number one. There are two possible structures $\de\,f\,\lambda\,\vepsilon\,(\diff x)^4$ and $f\,\de\,\lambda\,\,\vepsilon\,(\diff x)^4$, but they bring to the same integrated anomaly, as it follows by integrating by parts. Therefore, we consider without loss of generality the second structure. In writing down all the possible contractions, it is easier to consider the dual $f\,\de\,\lambda\,\diff^4 x$, where $\diff^4x$ is the measure of integration in four dimensions. It remains to contract the four indices of $f\,\de\,\lambda$ in all the possible ways allowed by the symmetries of $f$. The result is a unique possibility, $f_{\mu\nu}{}^\nu\,\de^\mu\,\lambda\,\diff^4 x$. But this is precisely the anomaly found before. Indeed, if $n=2$, the formula \eqref{Anomaly2nd} reads
\begin{align}
\omega^{(4)}_1 &= f^{(2)}_\mu\,\de_\nu\,\lambda\,{\star(\diff x^\mu\,\diff x^\nu)} = \nn\\
& = \tfrac{1}{2}\,f^{(2)}_\mu\,\de_\nu\lambda\,\vepsilon^{\mu\nu}{}_{\alpha\beta}\,\diff x^\alpha\,\diff x^\beta.
\end{align}
whose dual is precisely proportional to $f_{\mu\nu}{}^\nu\,\de^\mu\,\lambda\,\diff^4 x$.

\subsection{More solutions in six dimensions}\label{More6d}

In six dimensions the \textsc{brst} cohomology is nontrivial only on the five-forms of ghost number two and on the six-forms of ghost number one, so that the descent is:
\begin{equations}
& s\,\omega^{(6)}_1 = -\,\diff\omega^{(5)}_2, \\
& s\,\omega^{(5)}_2 = 0.
\end{equations}
The solution found in the previous section reads
\begin{equation}
\omega_7 = f^{(2)}_\mu\,H_\nu\,H_\vrho\,{\star(\diff x^\mu\,\diff x^\nu\,\diff x^\vrho)},
\end{equation}
whose components are
\begin{equations}
\omega_1^{(6)} &= 2\,f^{(2)}_\mu\,h^{(1)}_\nu\,\de_\vrho\,\lambda\,{\star(\diff x^\mu\,\diff x^\nu\,\diff x^\vrho)},\\
\omega_2^{(5)} &= 2\,f^{(2)}_\mu\,\de_\nu\,\lambda\,\de_\vrho\,\lambda\,{\star(\diff x^\mu\,\diff x^\nu\,\diff x^\vrho)}.
\end{equations}
This solution is not unique, since many other contractions, involving the form indices of $f$, are available. Such solutions are not well-defined from the polyform point of view, in the sense that $\delta$ is defined to act on $f$ only if it appears as a two-form $f^{(2)}_\mu$. If one wants to compute the $\delta$ variation of the components $f_{\alpha\beta\mu}$, it is necessary to split $\delta = s + \diff$. Namely,
\begin{equation}
\delta\,f_{\alpha\beta\mu} = (s + \diff)\,f_{\alpha\beta\mu} = \diff x^\beta\,\de_\beta\,f_{\alpha\beta\mu} .
\end{equation} 
Nevertheless, we can use polyforms also in this case as a trick to verify more easily that the descent is solved by an expression. For example, consider the following polyform:
\begin{align}
\tilde{\omega}_7 &= \de_\mu\,f^{(2)}_\nu\,\lambda\,H_\vrho\,{\star(\diff x^\mu\,\diff x^\nu\,\diff x^\vrho)} + \diff x^\alpha\,f_{\alpha\mu\nu}\,\diff x^\beta\,H_\beta\,H_\vrho\,{\star(\diff x^\mu\,\diff x^\nu\,\diff x^\vrho)} = \nn\\
& =  (\de_\mu\,f^{(2)}_\nu\,\lambda + \diff x^\alpha\,f_{\alpha\mu\nu}\,\diff\lambda)\,H_\vrho\,{\star(\diff x^\mu\,\diff x^\nu\,\diff x^\vrho)}.
\end{align}
It solves the Stora-Zumino equation. Indeed, 
\begin{align}
\delta\,\tilde{\omega}_7 &= (\de_\mu\,f^{(2)}_\nu\,\diff\lambda - \diff x^\alpha\,\delta\,f_{\alpha\mu\nu}\,\diff\lambda)\,H_\vrho\,{\star(\diff x^\mu\,\diff x^\nu\,\diff x^\vrho)} \,+\nn\\
& - (\de_\mu\,f^{(2)}_\nu\,\lambda + \diff x^\alpha\,f_{\alpha\mu\nu}\,\diff\lambda)\,f^{(2)}_\vrho\,{\star(\diff x^\mu\,\diff x^\nu\,\diff x^\vrho)}.
\end{align}
The second line vanishes in six dimensions, being a seven-form. Replacing $\delta\,f_{\alpha\mu\nu}$ in the first line,
\begin{align}
\delta\,\tilde{\omega}_7 &= \tfrac{1}{2}\,\diff x^\alpha\,\diff x^\beta\,(\de_\mu\,f_{\alpha\beta\nu} - \de_{[\beta}\,f_{\alpha]\mu\nu})\,\diff\lambda\,H_\vrho\,{\star(\diff x^\mu\,\diff x^\nu\,\diff x^\vrho)} = \nn\\
& = \tfrac{1}{2}\,\diff x^\alpha\,\diff x^\beta\,(\de_\mu\,f_{\alpha\beta\nu} + \de_{\alpha}\,f_{\beta\mu\nu} + \de_{\beta}\,f_{\mu\alpha\nu})\,\diff\lambda\,H_\vrho\,{\star(\diff x^\mu\,\diff x^\nu\,\diff x^\vrho)} = 0,
\end{align}
which vanishes thanks to the Bianchi identity $\de_{[\alpha}\,f_{\beta\mu]\nu}=0$. Therefore, we proved that
\begin{align}
\tilde{\omega}^{(6)}_1 &= (\de_\mu\,f^{(2)}_\nu\,\lambda + \diff x^\alpha\,f_{\alpha\mu\nu}\,\diff\lambda)\,h^{(1)}_\vrho\,{\star(\diff x^\mu\,\diff x^\nu\,\diff x^\vrho)} = \nn\\
& = \tfrac{1}{6}\,(\de_\mu\,f^{(2)}_\nu\,\lambda + \diff x^\sigma\,f_{\sigma\mu\nu}\,\diff\lambda)\,h^{(1)}_\vrho\,
\vepsilon^{\mu\nu\vrho}{}_{\alpha\beta\gamma}\,\diff x^\alpha\,\diff x^\beta\,\diff x^\gamma
\end{align}
is an anomaly in six dimensions.

Let us conclude by determining the cohomology on the six-forms of ghost number one. We know that there are two possible structures with the right dimensions, $f\,f\,\lambda\,\vepsilon\,(\diff x)^6$ and $\de\,f\,\de\,\lambda\,\vepsilon\,(\diff x)^6$. Actually, the second one has to be discharged, since the \emph{integrated} anomaly would be trivial. Indeed, by integrating by parts,
\begin{align}
\int \de\,f\,\de\,\lambda\,\vepsilon\,(\diff x)^6 &= -\int f\,\de\,\de\,\lambda\,\vepsilon\,(\diff x)^6 = \nn\\
& = -\int f\,s\,h\,\vepsilon\,(\diff x)^6 = 
-\,s\int f\,h\,\vepsilon\,(\diff x)^6,
\end{align}
for every contraction of the indices. Therefore, we consider only $f\,f\,\lambda\,\vepsilon\,(\diff x)^6$, or equivalently its dual $f\,f\,\lambda\,\diff^6 x$ where $\diff^6 x$ is the volume form. In principle, there are three possible contractions of two $f$'s: $f_{\mu\nu\vrho}\,f^{\mu\nu\vrho}$, $f_{\mu\nu\vrho}\,f^{\mu\vrho\nu}$ and $f_{\mu\nu}{}^\nu\,f^{\mu}{}_{\vrho}{}^{\vrho}$. Actually, the first or the second can be dropped out without loss of generality, as a consequence of the cyclicity property $f_{[\mu\nu\vrho]}=0$. Indeed, the following identity holds:
\begin{equation}
f_{\mu\nu\vrho}\,f^{\mu\vrho\nu} = \tfrac{1}{2}\,f_{\mu\nu\vrho}\,f^{\mu\nu\vrho} - \tfrac{1}{2}\,f_{[\mu\nu\vrho]}\,f^{\mu\nu\vrho}.
\end{equation}
Therefore, the most general expression in the fracton \textsc{brst} cohomology on the six-forms of ghost number one is 
\begin{equation}
{\omega^{\natural}}^{(6)}_1 = \alpha\,f_{\mu\nu\vrho}\,f^{\mu\nu\vrho}\,\lambda\,\diff^6 x + \beta\,f_{\mu\nu}{}^\nu\,f^\mu{}_\vrho{}^\vrho\,\lambda\,\diff^6 x,
\end{equation}
for any constant $\alpha$ and $\beta$.

\subsection{Comparison with type A Weyl anomalies}\label{Weyl}

In Section \ref{SomeSolutions} we showed that the cohomology class \eqref{EvenCohoClass} produces a solution of the descent only in two dimensions. It is interesting to observe that the cohomology class \eqref{EvenCohoClass} is the same as in computing the type \textsc{a} Weyl anomaly in conformal gravity in even spacetime dimensions. Consider a $2\,n$-dimensional manifold with metric tensor $g_{\mu\nu}$. The \textsc{brst} rules for the Weyl scaling of the metric and the Weyl ghost $\sigma$ (anticommutig scalar field) are
\begin{equations}
& s_w\,g_{\mu\nu} = 2\,\sigma\,g_{\mu\nu},\\
& s_w\,\sigma = 0.
\end{equations}
$s_w$ is obviously nilpotent on the space generated by $g_{\mu\nu}$, $\sigma$ and their derivatives.\footnote{$s_w$ can be thought as the equivariant \textsc{brst} operator in a diffeomorphism invariant theory (conformal gravity) when it acts on covariant tensor $s_w = s - \mathcal{L}_\xi$, where $\mathcal{L}_\xi$ is the Lie derivative, generating the infinitesimal diffeomorphisms, and the vector field $\xi$ is the ghost of diffeomorphisms \cite{Bonora:1985cq}.} The $s_w$ cohomology contains the following anomaly
\begin{equation}\label{WeylAnomaly}
E_1^{(2n)} = \sigma\,\vepsilon^{\mu_1\nu_1\dots\mu_n\nu_n}\,R_{\mu_1\nu_1}\dots R_{\mu_n\nu_n},
\end{equation}
which is the famous type \textsc{a} Weyl anomaly in conformal gravity, $R_{\mu\nu}$ being the Riemann two-form $R_{\mu\nu} := \tfrac{1}{2}\,R_{\alpha\beta\mu\nu}\,\diff x^\alpha\,\diff x^\beta$ \cite{Deser:1993yx, Imbimbo:1999bj}.
The starting point in the descent is 
\begin{equation}\label{StartingDescent}
E_n^{(n)} = \sigma\,{\star(\diff\sigma\dots\diff\sigma)},
\end{equation}
which is precisely the same as \eqref{EvenCohoClass}, by replacing $\sigma \leftrightarrow \lambda$, as anticipated. One can show quickly that \eqref{StartingDescent} is the starting point of the descent, ending with \eqref{WeylAnomaly}, by using the polyform approach \cite{Imbimbo:Unpublished} (see the Appendix \ref{AppendixC1}). 

The link with the fracton cohomology is the following. Consider the  transformation of the \emph{linearized} $R_{\mu\nu}$, which reads\footnote{Indeed, if $g_{\mu\nu} = \eta_{\mu\nu} + \tilde{h}_{\mu\nu}$, then $s_w\,\tilde{h}_{\mu\nu} = 2\,\sigma\,\eta_{\mu\nu}$, or equivalently $s_w\,\tilde{h}^{(1)}_\mu = 2\,\sigma\,\diff x_\mu$, having defined $\tilde{h}^{(1)}_\mu = \tilde{h}_{\mu\alpha}\,\diff x^\alpha$. Moreover, $R_{\mu\nu} = \frac{1}{2}\,\de_{[\mu}\,\diff \tilde{h}^{(1)}_{\nu]}$, so that $s_w\,R_{\mu\nu} =  -\frac{1}{2}\,\de_{[\mu}\,\diff s_w\,\tilde{h}^{(1)}_{\nu]} = -\,\de_{[\mu}\,\diff\sigma\,\diff x_{\nu]}$.}
\begin{equation}
s_w\,R_{\mu\nu} = -\,\de_{[\mu}\,\diff \sigma\,\diff x_{\nu]}.
\end{equation}
Now, using the fracton \textsc{brst} rules \eqref{BRSTrule1}--\eqref{BRSTrule2}, one sees that
\begin{equation}
s\,h^{(1)}_{[\mu}\,\diff x_{\nu]} = -\,\de_{[\mu}\,\diff \lambda\,\diff x_{\nu]},
\end{equation}
which is precisely the $s_w$ variation of the linearized Riemann two-form, identifying
$R_{\mu\nu} \leftrightarrow h^{(1)}_{[\mu}\,\diff x_{\nu]}$ and $\sigma \leftrightarrow \lambda$. Indeed, using this identification, the two-dimensional fracton anomaly \eqref{Anomaly2d} is the type \textsc{a} Weyl anomaly. So, why not conclude that the type \textsc{a} Weyl anomaly \eqref{WeylAnomaly} is a fracton anomaly in $2\,n$ dimensions, upon replacing the Weyl ghost $\sigma$ with the fracton ghost $\lambda$ and the Riemann two-form $R^{\mu\nu}$ with the combination $h^{(1)}_{[\mu}\,\diff x_{\nu]}$? The reason why the type \textsc{a} Weyl anomaly is not a fracton anomaly is that the Riemann two-form satisfies a Bianchi identity, which in the linearized theory says that it is $\diff$-closed $\diff\,R^{\mu\nu} = 0$. The Bianchi identity is crucial in proving that \eqref{WeylAnomaly} is an anomaly, as shown in Appendix \ref{AppendixC1}. But $h^{(1)}_{[\mu}\,\diff x_{\nu]}$ does not satisfy this property, since its external differential is $f^{(2)}_{[\mu}\,\diff x_{\nu]}$, which is not vanishing. Instead, in two dimensions there is no such an obstruction, since the external differential of $h^{(1)}_{[\mu}\,\diff x_{\nu]}$ trivially vanishes -- it would be a three-form.

\section{An example of fracton matter theory}\label{Matter}

In this last section we study a possible matter theory to be covariantly coupled with the fracton gauge field, by starting with the models in \cite{Seiberg:2020bhn, Burnell:2021reh}.\footnote{Another possibility for a fracton current is considered in \cite{Afxonidis:2023pdq}. The author thanks E. Afxonidis, A. Caddeo, C. Hoyos and D. Musso for making him aware of it.} Consider a real scalar field $\vphi$ with mass dimension $[\vphi]=\frac{d-2}{2}$ in such a way that the kinetic term $\de\,\vphi\,\de\,\vphi$ depends on no mass scale. Introduce a scaled fracton gauge field $h^{(\mu)}_{\mu\nu}$, where $\mu$ is a mass scale ($[\mu]=1$), such that the kinetic term $\sim \de\,h^{(\mu)}\,\de\,h^{(\mu)}$ does not explicitly depend on the mass scale. This implies $[h^{(\mu)}_{\mu\nu}] = \frac{d-2}{2}$, and so we can define
\begin{equation}
h^{(\mu)}_{\mu\nu} = \mu^{\frac{d}{2}-3}\,h_{\mu\nu}.
\end{equation}
We require the coupling between $h^{(\mu)}_{\mu\nu}$ and the Noether current $J^{\mu\nu}$ to be minimal, as in \eqref{MinimalCoupling}. We can argue that $J \sim \de^m\,\vphi^n$, $m, n$ positive integers, with the derivatives arranged in all the possible ways. The integers $m,n$ should satisfy $m\geqslant 2$, since we need two free indices, and obviously $n \geqslant 1$. Since $[h_{\mu\nu}^{(\mu)}\,J^{\mu\nu}]=d$, $J^{\mu\nu}$ should have mass dimension $[J^{\mu\nu}]=\frac{d+2}{2}$. The unique possibility for each $d$ is $m=2$, $n=1$, so that the current reads
\begin{equation}\label{NoetherCurrent}
J_{\mu\nu} = \de_\mu\,\de_\nu\,\vphi.
\end{equation}

The simplest theory with \eqref{NoetherCurrent} as a Noether current is defined by the following higher derivative action:
\begin{equation}\label{MatterTheory}
S = \frac{1}{2\,\mu^2}\int \diff^d x\,\de_\mu\,\de_\nu\,\vphi\,\de^\mu\,\de^\nu\,\vphi,
\end{equation}
Indeed, the action is invariant under the global shift
\begin{equation}\label{GlobalSymmetry}
\delta\,\vphi = \mu^2\,\lambda^{(\mu)},
\end{equation}
with $\lambda^{(\mu)}$ being a constant with mass dimension $[\lambda^{(\mu)}]=\frac{d}{2}-3$, and the associated Noether current is the desired one
\begin{equation}
\lambda^{(\mu)}\,J^{\mu\nu} = \frac{\delta\,S}{\delta\,\de_\mu\,\de_\nu\,\vphi}\,\delta\,\vphi  = \lambda^{(\mu)}\,\de_\mu\,\de_\nu\,\vphi.
\end{equation}

The theory defined by the action \eqref{MatterTheory} involves higher derivatives. Usually, higher derivative terms are comprised in low energy effective theories. For example, string theory dictates higher derivative corrections to supergravity. In general, such terms improve the renormalizability, but at a classical level they are commonly believed to bring to unbounded energy spectra, and so to instabilities, as studied by Ostrogradsky long ago \cite{Ostrogradsky:1850fid}. Such instabilities, often called ``Ostrogradsky ghosts", in general persist at a quantum level too (see \cite{Woodard:2015zca} for an overview).

Higher derivative terms are often considered in models of modified gravity, such as massive gravity or Galileon gravity. In both the cases, an appropriate fine tuning of the coefficients in the Lagrangian prevents ghosts to rise. Although higher derivative terms cannot be removed from the Lagrangian by integrating by part, the equations of motion are still of the second order.

The simplest higher derivative model is the Pais-Uhlenbeck oscillator \cite{Pais:1950za}, whose Hamiltonian is equivalent to the difference of the Hamiltonian of two harmonic oscillators. So, it is not positive definite. At a quantum level, as shown in \cite{Gibbons:2019lmj}, it is possible to switch from a Fock space with positive definite norm states only, but with unbounded energy spectrum, to bounded energy spectrum, but indefinite norm in the Fock space. 

The theory defined by \eqref{MatterTheory} corresponds to the field theory associated to the Pais-Uhlenbeck oscillator with vanishing masses (also known as biharmonic scalar field theory). According to the Ostrogradsky formalism, the Hamiltonian density is defined by the Legendre transform $\mathcal{H} = \pi_1\,\dot{\vphi} + \pi_2\,\ddot{\vphi} - \mathcal{L}$, where two canonical momenta are defined, $\pi_1 = \frac{\de \mathcal{L}}{\de \dot{\vphi}}-\de_i\,\frac{\de \mathcal{L}}{\de \de_i\,\dot{\vphi}}-\de_t\,\frac{\de \mathcal{L}}{\de \ddot{\vphi}}$ and $\pi_2 = \frac{\de \mathcal{L}}{\de \ddot{\vphi}}$. Since the Lagrangian density of \eqref{MatterTheory} reads
\begin{equation}
\mathcal{L} = \frac{1}{2\,\mu^2}\,(\ddot{\vphi}^2 - 2\,\vec{\nabla}\,\dot{\vphi}\cdot\vec{\nabla}\,\dot{\vphi} + (\nabla^2\,\vphi)^2),
\end{equation}
using the convention $(+,-,\dots,-)$ for the signature of the metric, the corresponding Hamiltonian density is equal to
\begin{equation}
\mathcal{H} = \frac{1}{2\,\mu^2}\,(\ddot{\vphi}^2 - 2\,\vec{\nabla}\,\dot{\vphi}\cdot\vec{\nabla}\,\dot{\vphi} - (\nabla^2\,\vphi)^2 - 2\,\dot{\vphi}\,\dddot{\vphi}),
\end{equation}
which is not positive definite. Nevertheless, as argued in \cite{Smilga:2004cy}, there is a threshold under which the modes are stable. Instead, the vacuum is unstable with respect to small fluctuations. 

The equation of motion of the field $\vphi$ is the square of the the Laplace equation (biharmonic equation):
\begin{equation}
(\de^2)^2\,\vphi = 0.
\end{equation}
To solve it, one can set
\begin{equation}
\chi := \de^2\,\vphi.
\end{equation}
Then, the equation of motion imposes $\chi$ to be harmonic, or a massless scalar field, which can be expanded in Fourier modes as usual:
\begin{equation}
\chi(x) = \int\frac{\diff^{d-1} p}{2\,\omega_{\vec{p}}}\,e^{-ip\cdot x}\,a_+(\vec{p}) + e^{i\,p\cdot x}\,a_-(\vec{p}),
\end{equation}
where $\omega_{\vec{p}}^2 = |\vec{p}|^2$ and $a_\pm(\vec{p})=a(p^0=\omega_{\vec{p}},\vec{p})$, $a$ being an arbitrary function. To find $\vphi$ one has to solve the Laplace equation with an (harmonic) source. The most general solution is the most general solution of the homogeneous equation, so again a harmonic function $\vphi_0$, plus a particular solution of the sourced equation. The latter can be found by computing the Green function $G(x,x')$ of the square Laplacian operator, given suitable boundary conditions:
\begin{equation}
\vphi(x) = \vphi_0(x) + \int \diff x'\,G(x,x')\,\chi(x').
\end{equation} 

Now, let us study how to gauge the shift global symmetry \eqref{GlobalSymmetry}, that is, how to get a theory invariant under a spacetime dependent shift:
\begin{equation}
\delta_\lambda\,\vphi(x) = \mu^2\,\lambda^{(\mu)}(x).
\end{equation} 
To get the improved action $\hat{S}$ we can simply replace 
\begin{equation}
\de_\mu\,\de_\nu\,\vphi \rightarrow \de_\mu\,\de_\nu\,\vphi - \mu^2\,h^{(\mu)}_{\mu\nu},
\end{equation}
where $h^{(\mu)}_{\mu\nu}$ is the (rescaled) fracton gauge field. Indeed, provided that the gauge parameter is precisely the local shift $\lambda^{(\mu)}(x)$\footnote{The relation between $\lambda^{(\mu)}$ and $\lambda$ ($[\lambda]=0$) is $\lambda^{(\mu)} = \mu^{\frac{d}{2}-3}\,\lambda$.}
\begin{equation}
\delta_\lambda\,h_{\mu\nu}^{(\mu)}(x) = \de_\mu\,\de_\nu\,\lambda^{(\mu)}(x),
\end{equation} the shifted expression is manifestly gauge invariant. The improved action reads
\begin{align}
\hat{S} &=  \int\diff^d x\,\frac{1}{2\,\mu^2}\,(\de_\mu\,\de_\nu\,\vphi - \mu^2\,h^{(\mu)}_{\mu\nu})\,(\de^\mu\,\de^\nu\,\vphi - \mu^2\,h^{(\mu)\mu\nu}) = \nn\\
&= \int\diff^d x\;\,
\frac{1}{2\,\mu^2}\,\de_\mu\,\de_\nu\,\vphi\,\de^\mu\,\de^\nu\,\vphi - h^{(\mu)}_{\mu\nu}\,J^{\mu\nu} + \frac{1}{2}\,\mu^2\,h^{(\mu)}_{\mu\nu}\,h^{(\mu)\mu\nu}.
\end{align}
Besides the minimal coupling term, there is also a massive term for $h^{(\mu)}_{\mu\nu}$, with mass $\mu$. Thus, if a gauge invariant kinetic term for $h^{(\mu)}_{\mu\nu}$ is included, $\vphi$ can be viewed as the St\"uckelberg field one introduces in order to restore the gauge symmetry, when the gauge breaking mass term is added to the kinetic action \cite{Stueckelberg:1938hvi}. 

The theory defined by the action \eqref{MatterTheory} is an attempt to make covariant some theories studied in the condensed matter context, where they appear as low energy continuum limit of lattice models, such as the \textsc{xy} model (in the three-dimensional case)  (see \cite{Seiberg:2020bhn, Burnell:2021reh} and the reference within).\footnote{In particular, the proposed matter action (if $d=3$) is the covariant version of the \emph{plaquette} \textsc{xy}-\emph{model} \cite{Seiberg:2020bhn}. See also \cite{Brauner:2022rvf} for a quick summary of the existing literature.} In this context fracton excitations emerged for the first time. It was shown that all the features of the fracton electromagnetism can be captured by using the covariant formulation \cite{Bertolini:2022ijb}. One may expect that it is possible to exhibit a covariant matter theory mimicking the features of the fracton matter. It would be interesting to investigate if the action \eqref{MatterTheory}, or a modification of it, is the low energy continuum limit of a lattice model exhibiting fracton excitations, in the same way as, for example, antiferromagnetic chains were shown to exhibit dyon excitations in the continuum limit \cite{Affleck:1986}.

\section{Summary and outlook}

Studying the \textsc{brst} fracton cohomology mod $\diff$, defined by the rules \eqref{BRSTrule1}--\eqref{BRSTrule2}, we showed the following 
\begin{itemize}
\item[--] There is no anomaly in odd spacetime dimensions.
\item[--] The most general anomaly in two dimensions is
\begin{equation}
\omega^{(2)}_1 = \lambda\,\vepsilon^{\mu\nu}\,h^{(1)}_\mu\,\diff x_\nu.
\end{equation}
\item[--] The most general anomaly in four dimensions is 
\begin{equation}
\omega^{(4)}_1 = \tfrac{1}{2}\,f^{(2)}_\mu\,\de_\nu\lambda\,\vepsilon^{\mu\nu}{}_{\alpha\beta}\,\diff x^\alpha\,\diff x^\beta.
\end{equation}
\item[--] The following are anomalies in six dimensions
\begin{align}
\omega_1^{(6)} &= \tfrac{1}{3}\,f^{(2)}_\mu\,h^{(1)}_\nu\,\de_\vrho\,\lambda\,\vepsilon^{\mu\nu\vrho}{}_{\alpha\beta\gamma}\,\diff x^\alpha\,\diff x^\beta\,\diff x^\gamma,\\
\tilde{\omega}^{(6)}_1 &= \tfrac{1}{6}\,(\de_\mu\,f^{(2)}_\nu\,\lambda + \diff x^\sigma\,f_{\sigma\mu\nu}\,\diff\lambda)\,h^{(1)}_\vrho\,
\vepsilon^{\mu\nu\vrho}{}_{\alpha\beta\gamma}\,\diff x^\alpha\,\diff x^\beta\,\diff x^\gamma,\\
{\omega^{\natural}}^{(6)}_1 &= f_{\mu\nu\vrho}\,f^{\mu\nu\vrho}\,\lambda\,\diff^6 x + \alpha\,f_{\mu\nu}{}^\nu\,f^\mu{}_\vrho{}^\vrho\,\lambda\,\diff^6 x,
\end{align}
for arbitrary constant $\alpha$. 
\item[--] The following is an anomaly in $2\,n$ dimensions
\begin{align}
\omega_1^{(2n)} &= \tfrac{1}{n(n-2)!}\,f^{(2)}_{\mu_1}\,h_{\mu_2}^{(1)}\dots h_{\mu_{n-1}}^{(1)}\,\de_{\mu_n}\,\lambda\,\times\nn\\
& \times\vepsilon^{\mu_1\mu_2\dots\mu_{n-1}\mu_n}{}_{\alpha_1\alpha_2\dots\alpha_{n-1}\alpha_n}\,\diff x^{\alpha_1}\,\diff x^{\alpha_2}\dots\diff x^{\alpha_{n-1}}\diff x^{\alpha_n}.
\end{align}
\end{itemize}
In this paper we addressed the issue of classifying the cohomology classes in fracton theories, which are the model-independent part of anomalies. Whether they constitute anomalies in a physical theory depends on the specific models that are considered. The St\"uckelberg-like model discussed in Section \ref{Matter} does not required to be regularized. Indeed the symmetry transformation on the matter field is a simple shift: For instance, the measure of integration in the path integral is invariant, and so, according to the Fujikawa approach \cite{Fujikawa:1979ay}, there is no anomaly. 

Moreover, it should be kept in mind that possible fracton models derived from lattice theories should not be afflicted by anomalies, by means of the Nielsen-Ninomiya no-go theorem \cite{Nielsen:1980rz, Nielsen:1981xu}, which prevents a net chirality in a lattice model of fermions, if the Hamiltonian of the theory is quadratic in the fields, phase and translation invariant and local.\footnote{The author thanks the referee to have pointed this out.} Therefore, it remains to find other possible fracton matter theories (or investigate according to this point of view the matter theory introduced in \cite{Afxonidis:2023pdq}) and to test if the fracton symmetry is anomalous or not in such theories.

\section*{Acknowledgments}

The author thanks Daniel Sacco Shaikh for his collaboration at the early stages of this project and for useful and stimulating discussions, and Prof. Camillo Imbimbo for useful discussions and, in particular, for having shared some observations on the $\delta$ cohomology.

\appendix
\section{Appendix. Polyform proof of type A Weyl anomaly}\label{AppendixC1}

In the following we review the proof that the type \textsc{a} Weyl anomaly \eqref{WeylAnomaly} is an anomaly in $2\,n$ dimensions, using the polyform approach, as shown in \cite{Imbimbo:Unpublished}.\footnote{A similar approach can be found in \cite{Boulanger:2007st, Boulanger:2007ab}.} Notice that the expression
\begin{equation}
s\,R_{\mu\nu} = - \,\de_{[\mu}\,\diff \sigma\,\diff x_{\nu]},
\end{equation}
can be equivalently written as
\begin{equation}\label{deltaInvariance}
\delta\,(R_{\mu\nu} + \de_{[\mu}\,\sigma\,\diff x_{\nu]}) = 0,
\end{equation}
since $R^{\mu\nu}$ is $\diff$-closed (Bianchi identity) and $\de_{[\mu}\,\sigma\,\diff x_{\nu]}$ is $s$ invariant. Now consider
\begin{equation}
\omega_{2n+1} = \sigma\,\vepsilon^{\mu_1\nu_1\dots\mu_n\nu_n}\,(R_{\mu_1\nu_1}+ \de_{[\mu_1}\,\sigma\,\diff x_{\nu_1]})\dots (R_{\mu_n\nu_n}+\de_{[\mu_n}\,\sigma\,\diff x_{\nu_n]})
\end{equation}
whose component with ghost number one is the type \textsc{a} Weyl anomaly and the component with maximum ghost number is proportional to $\sigma\,{\star(\diff\sigma)^n}$. Using \eqref{deltaInvariance} and $\delta\,\sigma = \diff \sigma$, its $\delta$ variation is 
\begin{equation}
\delta\,\omega_{2n+1} = \diff\sigma\,\vepsilon^{\mu_1\nu_1\dots\mu_n\nu_n}\,(R_{\mu_1\nu_1}+ \de_{[\mu_1}\,\sigma\,\diff x_{\nu_1]})\dots (R_{\mu_n\nu_n}+\de_{[\mu_n}\,\sigma\,\diff x_{\nu_n]}),
\end{equation}
which reads also
\begin{align}
\delta\,\omega_{2n+1} &= \sum_{p=0}^n 2^{n-2p}\,\diff x^\mu\,\de_\mu\,\sigma\,\vepsilon^{\mu_1\nu_1\dots\mu_p\nu_p\mu_{p+1}\nu_{p+1}\dots\mu_n\nu_n}\,\times\\
& \qquad\;\;\times\,\diff x^{\alpha_1}\,\diff x^{\beta_1}\,R_{\alpha_1\beta_1\mu_1\nu_1}\dots \diff x^{\alpha_p}\,\diff x^{\beta_p}\,R_{\alpha_p\beta_p\mu_p\nu_p}\,\times\nn\\
& \qquad\;\;\times\,\de_{\mu_{p+1}}\,\sigma\,\diff x_{\nu_{p+1}}\dots \de_{\mu_n}\,\sigma\,\diff x_{\nu_n}.
\end{align}
Dualizing the general term in the sum, we can replace the set of differentials and the set of derivatives of $\sigma$ with two Levi-Civita tensors:
\begin{align}
& \vepsilon_{\mu\mu_{p+1}\dots\mu_n\vrho_1\dots\vrho_{n-1+p}}\,\vepsilon^{\mu_1\dots\mu_p\mu_{p+1}\dots\mu_n\nu_1\dots\nu_p\nu_{p+1}\dots\nu_n}\,\times\nn\\
\times\; & \,\vepsilon^{\mu\alpha_1\dots\alpha_p\beta_1\dots\beta_p}{}_{\nu_{p+1}\dots\nu_n\gamma_1\dots\gamma_{n-1-p}}\,R_{\alpha_1\beta_1\mu_1\nu_1}\dots R_{\alpha_p\beta_p\mu_p\nu_p}.
\end{align}
Contracting the first two $\vepsilon$, one gets, up to a constant,
\begin{equation}
\delta^{\mu_1\dots\mu_p\nu_1\dots\nu_p\,\nu_{p+1}\dots\nu_n}_{\mu\vrho_1\dots\vrho_{n-1+p}}\,\vepsilon^{\mu\alpha_1\dots\alpha_p\beta_1\dots\beta_p\nu_{p+1}\dots\nu_n}{}_{\gamma_1\dots\gamma_{n-1-p}}\,R_{\alpha_1\beta_1\mu_1\nu_1}\dots R_{\alpha_p\beta_p\mu_p\nu_p}.
\end{equation}
Now, if $\mu = \mu_i$ or if $\mu = \nu_i$, for any $i$ between 1 and $p$, then it appears the antisymmetrization $R_{[\alpha_i\beta_i\mu]\nu_i}$ or $R_{[\alpha_i\beta_i\mu]\mu_i}$, which vanishes thanks to the Bianchi identity. If $\mu = \nu_i$, for any $i$ between $p+1$ and $n$, then two $\mu$ saturated in the Levi-Civita tensor appear, so the expression vanishes. Therefore, we conclude that 
\begin{equation}
\delta\,\omega_{2n+1} = 0,
\end{equation}
as we wanted to show. 

This proof regards the linearized theory, but it can be readily extended to the general case. Indeed, the $s$ variation of the whole Riemann two-form and the Bianchi identity are the same as before, but with the simple derivative replaced by the gravitational covariant derivative (with the Levi-Civita connection). Since the Riemann two-form appears in a scalar contraction in $\omega_{2n+1}$, the proof analogously follows.

\bibliographystyle{JHEP}
\bibliography{ir}

\providecommand{\href}[2]{#2}\begingroup\raggedright\begin{thebibliography}{10}

\bibitem{Pretko:2017xar}
M.~Pretko, \emph{{Higher-Spin Witten Effect and Two-Dimensional Fracton
  Phases}}, \href{https://doi.org/10.1103/PhysRevB.96.125151}{\emph{Phys. Rev.
  B} {\bfseries 96} (2017) 125151}
  [\href{https://arxiv.org/abs/1707.03838}{{\ttfamily 1707.03838}}].

\bibitem{Seiberg:2020bhn}
N.~Seiberg and S.-H.~Shao, \emph{{Exotic Symmetries, Duality, and Fractons in
  2+1-Dimensional Quantum Field Theory}},
  \href{https://doi.org/10.21468/SciPostPhys.10.2.027}{\emph{SciPost Phys.}
  {\bfseries 10} (2021) 027}
  [\href{https://arxiv.org/abs/2003.10466}{{\ttfamily 2003.10466}}].

\bibitem{Burnell:2021reh}
F.J.~Burnell, T.~Devakul, P.~Gorantla, H.T.~Lam and S.-H.~Shao, \emph{{Anomaly
  inflow for subsystem symmetries}},
  \href{https://doi.org/10.1103/PhysRevB.106.085113}{\emph{Phys. Rev. B}
  {\bfseries 106} (2022) 085113}
  [\href{https://arxiv.org/abs/2110.09529}{{\ttfamily 2110.09529}}].

\bibitem{Alexander:1982}
S.~Alexander and R.~Orbach, \emph{{Density of states on fractals:
  ``fractons"}}, {\emph{Jour. Phys. Lett.} {\bfseries 43} (1982) 625}.

\bibitem{Vijay:2015mka}
S.~Vijay, J.~Haah and L.~Fu, \emph{{A New Kind of Topological Quantum Order: A
  Dimensional Hierarchy of Quasiparticles Built from Stationary Excitations}},
  \href{https://doi.org/10.1103/PhysRevB.92.235136}{\emph{Phys. Rev. B}
  {\bfseries 92} (2015) 235136}
  [\href{https://arxiv.org/abs/1505.02576}{{\ttfamily 1505.02576}}].

\bibitem{Khlopov:1981wm}
M.Y.~Khlopov, \emph{{Fractionally charged particles and quark confinement}},
  {\emph{Pisma Zh. Eksp. Teor. Fiz.} {\bfseries 33} (1981) 170}.

\bibitem{Blasi:2022mbl}
A.~Blasi and N.~Maggiore, \emph{{The theory of symmetric tensor field: From
  fractons to gravitons and back}},
  \href{https://doi.org/10.1016/j.physletb.2022.137304}{\emph{Phys. Lett. B}
  {\bfseries 833} (2022) 137304}
  [\href{https://arxiv.org/abs/2207.05956}{{\ttfamily 2207.05956}}].

\bibitem{Bertolini:2022ijb}
E.~Bertolini and N.~Maggiore, \emph{{Maxwell theory of fractons}},
  \href{https://doi.org/10.1103/PhysRevD.106.125008}{\emph{Phys. Rev. D}
  {\bfseries 106} (2022) 125008}
  [\href{https://arxiv.org/abs/2209.01485}{{\ttfamily 2209.01485}}].

\bibitem{Bertolini:2023juh}
E.~Bertolini, A.~Blasi, A.~Damonte and N.~Maggiore, \emph{{Gauging Fractons and
  Linearized Gravity}},
  \href{https://doi.org/10.3390/sym15040945}{\emph{Symmetry} {\bfseries 15}
  (2023) 945} [\href{https://arxiv.org/abs/2304.10789}{{\ttfamily
  2304.10789}}].

\bibitem{Becchi:1974md}
C.~Becchi, A.~Rouet and R.~Stora, \emph{{Renormalization of the Abelian
  Higgs-Kibble Model}}, \href{https://doi.org/10.1007/BF01614158}{\emph{Commun.
  Math. Phys.} {\bfseries 42} (1975) 127}.

\bibitem{Becchi:1974xu}
C.~Becchi, A.~Rouet and R.~Stora, \emph{{The Abelian Higgs-Kibble Model.
  Unitarity of the S Operator}},
  \href{https://doi.org/10.1016/0370-2693(74)90058-6}{\emph{Phys. Lett. B}
  {\bfseries 52} (1974) 344}.

\bibitem{Becchi:1975nq}
C.~Becchi, A.~Rouet and R.~Stora, \emph{{Renormalization of Gauge Theories}},
  \href{https://doi.org/10.1016/0003-4916(76)90156-1}{\emph{Annals Phys.}
  {\bfseries 98} (1976) 287}.

\bibitem{Tyutin:1975qk}
I.V.~Tyutin, \emph{{Gauge Invariance in Field Theory and Statistical Physics in
  Operator Formalism}},  \href{https://arxiv.org/abs/0812.0580}{{\ttfamily
  0812.0580}}.

\bibitem{Stora:1976LM}
R.~Stora, \emph{Continuum gauge theories},  in \emph{New developments in
  quantum field theory and statistical mechanics, Cargese 1976, Ed. NATO ASI
  Ser B vol. 26}, P.M.~M.~Levy, ed., pp.~201--224, Plenum Press (1977),
  \href{https://doi.org/10.1007/978-1-4615-8918-1}{DOI}.

\bibitem{Stora:1984}
R.~Stora, \emph{Algebraic structure and topological origin of anomalies},  in
  \emph{Progress in Gauge Field Theory}, G.~'t~Hooft, A.~Jaffe, H.~Lehmann,
  P.K.~Mitter, I.M.~Singer and R.~Stora, eds., (Boston, MA), pp.~543--562,
  Springer US (1984), \href{https://doi.org/10.1007/978-1-4757-0280-4_19}{DOI}.

\bibitem{Zumino:1983ew}
B.~Zumino, \emph{{Chiral Anomalies and Differential Geometry: Lectures given at
  Les Houches, August 1983}},  in \emph{{Les Houches Summer School on
  Theoretical Physics: Relativity, Groups and Topology}}, pp.~1291--1322, 10,
  1983.

\bibitem{Manes:1985df}
J.~Manes, R.~Stora and B.~Zumino, \emph{{Algebraic Study of Chiral Anomalies}},
  \href{https://doi.org/10.1007/BF01208825}{\emph{Commun. Math. Phys.}
  {\bfseries 102} (1985) 157}.

\bibitem{Langouche:1984gn}
F.~Langouche, T.~Schucker and R.~Stora, \emph{{Gravitational Anomalies of the
  Adler-bardeen Type}},
  \href{https://doi.org/10.1016/0370-2693(84)90057-1}{\emph{Phys. Lett. B}
  {\bfseries 145} (1984) 342}.

\bibitem{Thierry-Mieg:1979fvq}
J.~Thierry-Mieg, \emph{{Geometrical reinterpretation of Faddeev-Popov ghost
  particles and BRS transformations}},
  \href{https://doi.org/10.1063/1.524385}{\emph{J. Math. Phys.} {\bfseries 21}
  (1980) 2834}.

\bibitem{Zumino:1983rz}
B.~Zumino, Y.-S.~Wu and A.~Zee, \emph{{Chiral Anomalies, Higher Dimensions, and
  Differential Geometry}},
  \href{https://doi.org/10.1016/0550-3213(84)90259-1}{\emph{Nucl. Phys. B}
  {\bfseries 239} (1984) 477}.

\bibitem{Imbimbo:2023sph}
C.~Imbimbo, D.~Rovere and A.~Warman, \emph{{Superconformal anomalies from
  superconformal Chern-Simons polynomials}},
  \href{https://doi.org/10.1007/JHEP05(2024)277}{\emph{JHEP} {\bfseries 05}
  (2024) 277} [\href{https://arxiv.org/abs/2311.05684}{{\ttfamily
  2311.05684}}].

\bibitem{Deser:1993yx}
S.~Deser and A.~Schwimmer, \emph{{Geometric classification of conformal
  anomalies in arbitrary dimensions}},
  \href{https://doi.org/10.1016/0370-2693(93)90934-A}{\emph{Phys. Lett. B}
  {\bfseries 309} (1993) 279}
  [\href{https://arxiv.org/abs/hep-th/9302047}{{\ttfamily hep-th/9302047}}].

\bibitem{Gromov:2018nbv}
A.~Gromov, \emph{{Towards classification of Fracton phases: the multipole
  algebra}}, \href{https://doi.org/10.1103/PhysRevX.9.031035}{\emph{Phys. Rev.
  X} {\bfseries 9} (2019) 031035}
  [\href{https://arxiv.org/abs/1812.05104}{{\ttfamily 1812.05104}}].

\bibitem{Wu:1988py}
Y.S.~Wu and A.~Zee, \emph{{Membranes, Higher Hopf Maps, and Phase
  Interactions}},
  \href{https://doi.org/10.1016/0370-2693(88)90882-9}{\emph{Phys. Lett. B}
  {\bfseries 207} (1988) 39}.

\bibitem{Brandt:1989rd}
F.~Brandt, N.~Dragon and M.~Kreuzer, \emph{{All consistent Yang-Mills
  anomalies}}, \href{https://doi.org/10.1016/0370-2693(89)90211-6}{\emph{Phys.
  Lett. B} {\bfseries 231} (1989) 263}.

\bibitem{Piguet:1995er}
O.~Piguet and S.P.~Sorella, \emph{{Algebraic renormalization: Perturbative
  renormalization, symmetries and anomalies}}, vol.~28, Springer Netherlands
  (1995),
  \href{https://doi.org/10.1007/978-3-540-49192-7}{10.1007/978-3-540-49192-7}.

\bibitem{Bonora:1985cq}
L.~Bonora, P.~Pasti and M.~Bregola, \emph{{Weyl Cocycles}},
  \href{https://doi.org/10.1088/0264-9381/3/4/018}{\emph{Class. Quant. Grav.}
  {\bfseries 3} (1986) 635}.

\bibitem{Imbimbo:1999bj}
C.~Imbimbo, A.~Schwimmer, S.~Theisen and S.~Yankielowicz,
  \emph{{Diffeomorphisms and holographic anomalies}},
  \href{https://doi.org/10.1088/0264-9381/17/5/322}{\emph{Class. Quant. Grav.}
  {\bfseries 17} (2000) 1129}
  [\href{https://arxiv.org/abs/hep-th/9910267}{{\ttfamily hep-th/9910267}}].

\bibitem{Imbimbo:Unpublished}
C.~Imbimbo and D.~Rovere, \emph{{Unpublished}}, .

\bibitem{Afxonidis:2023pdq}
E.~Afxonidis, A.~Caddeo, C.~Hoyos and D.~Musso, \emph{{Fracton gravity from
  spacetime dipole symmetry}},
  \href{https://doi.org/10.1103/PhysRevD.109.065013}{\emph{Phys. Rev. D}
  {\bfseries 109} (2024) 065013}
  [\href{https://arxiv.org/abs/2311.01818}{{\ttfamily 2311.01818}}].

\bibitem{Ostrogradsky:1850fid}
M.~Ostrogradsky, \emph{{M\'emoires sur les \'equations diff\'erentielles,
  relatives au probl\`eme des isop\'erim\`etres}}, {\emph{Mem. Acad. St.
  Petersbourg} {\bfseries 6} (1850) 385}.

\bibitem{Woodard:2015zca}
R.P.~Woodard, \emph{{Ostrogradsky's theorem on Hamiltonian instability}},
  \href{https://doi.org/10.4249/scholarpedia.32243}{\emph{Scholarpedia}
  {\bfseries 10} (2015) 32243}
  [\href{https://arxiv.org/abs/1506.02210}{{\ttfamily 1506.02210}}].

\bibitem{Pais:1950za}
A.~Pais and G.E.~Uhlenbeck, \emph{{On Field theories with nonlocalized
  action}}, \href{https://doi.org/10.1103/PhysRev.79.145}{\emph{Phys. Rev.}
  {\bfseries 79} (1950) 145}.

\bibitem{Gibbons:2019lmj}
G.W.~Gibbons, C.N.~Pope and S.~Solodukhin, \emph{{Higher Derivative Scalar
  Quantum Field Theory in Curved Spacetime}},
  \href{https://doi.org/10.1103/PhysRevD.100.105008}{\emph{Phys. Rev. D}
  {\bfseries 100} (2019) 105008}
  [\href{https://arxiv.org/abs/1907.03791}{{\ttfamily 1907.03791}}].

\bibitem{Smilga:2004cy}
A.V.~Smilga, \emph{{Benign versus malicious ghosts in higher-derivative
  theories}},
  \href{https://doi.org/10.1016/j.nuclphysb.2004.10.037}{\emph{Nucl. Phys. B}
  {\bfseries 706} (2005) 598}
  [\href{https://arxiv.org/abs/hep-th/0407231}{{\ttfamily hep-th/0407231}}].

\bibitem{Stueckelberg:1938hvi}
E.C.G.~Stueckelberg, \emph{{Interaction energy in electrodynamics and in the
  field theory of nuclear forces}},
  \href{https://doi.org/10.5169/seals-110852}{\emph{Helv. Phys. Acta}
  {\bfseries 11} (1938) 225}.

\bibitem{Brauner:2022rvf}
T.~Brauner, S.A.~Hartnoll, P.~Kovtun, H.~Liu, M.~Mezei, A.~Nicolis et~al.,
  \emph{{Snowmass White Paper: Effective Field Theories for Condensed Matter
  Systems}},  in \emph{{Snowmass 2021}}, 3, 2022
  [\href{https://arxiv.org/abs/2203.10110}{{\ttfamily 2203.10110}}].

\bibitem{Affleck:1986}
I.~Affleck, \emph{{Dyon Analogs in Antiferromagnetic Chains}}, {\emph{Phys.
  Rev. Lett.} {\bfseries 57} (1986) 1048}.

\bibitem{Fujikawa:1979ay}
K.~Fujikawa, \emph{{Path Integral Measure for Gauge Invariant Fermion
  Theories}}, \href{https://doi.org/10.1103/PhysRevLett.42.1195}{\emph{Phys.
  Rev. Lett.} {\bfseries 42} (1979) 1195}.

\bibitem{Nielsen:1980rz}
H.B.~Nielsen and M.~Ninomiya, \emph{{Absence of Neutrinos on a Lattice. 1.
  Proof by Homotopy Theory}},
  \href{https://doi.org/10.1016/0550-3213(82)90011-6}{\emph{Nucl. Phys. B}
  {\bfseries 185} (1981) 20}.

\bibitem{Nielsen:1981xu}
H.B.~Nielsen and M.~Ninomiya, \emph{{Absence of Neutrinos on a Lattice. 2.
  Intuitive Topological Proof}},
  \href{https://doi.org/10.1016/0550-3213(81)90524-1}{\emph{Nucl. Phys. B}
  {\bfseries 193} (1981) 173}.

\bibitem{Boulanger:2007st}
N.~Boulanger, \emph{{General solutions of the Wess-Zumino consistency condition
  for the Weyl anomalies}},
  \href{https://doi.org/10.1088/1126-6708/2007/07/069}{\emph{JHEP} {\bfseries
  07} (2007) 069} [\href{https://arxiv.org/abs/0704.2472}{{\ttfamily
  0704.2472}}].

\bibitem{Boulanger:2007ab}
N.~Boulanger, \emph{{Algebraic Classification of Weyl Anomalies in Arbitrary
  Dimensions}},
  \href{https://doi.org/10.1103/PhysRevLett.98.261302}{\emph{Phys. Rev. Lett.}
  {\bfseries 98} (2007) 261302}
  [\href{https://arxiv.org/abs/0706.0340}{{\ttfamily 0706.0340}}].

\end{thebibliography}\endgroup

\end{document}